\documentclass[sigconf]{acmart}

\usepackage[T1]{fontenc}
\usepackage{booktabs} 
\usepackage{multirow}
\usepackage{subcaption}
\usepackage{soul}

\usepackage{amsmath}
\newcommand{\mathbbm}[1]{\text{\usefont{U}{bbm}{m}{n}#1}} 

\setcopyright{rightsretained}

\copyrightyear{2018} 
\acmYear{2018} 
\setcopyright{acmcopyright}
\acmConference[SIGIR '18]{The 41st International ACM SIGIR Conference on Research \& Development in Information Retrieval}{July 8--12, 2018}{Ann Arbor, MI, USA}
\acmBooktitle{SIGIR '18: The 41st International ACM SIGIR Conference on Research \& Development in Information Retrieval, July 8--12, 2018, Ann Arbor, MI, USA}
\acmPrice{15.00}
\acmDOI{10.1145/3209978.3209991}
\acmISBN{978-1-4503-5657-2/18/07}

\begin{document}
\title{Collaborative Memory Network for Recommendation Systems}
\author{Travis Ebesu}
\affiliation{%
  \institution{Santa Clara University}
  \institution{Department of Computer Engineering}
  \city{Santa Clara}
  \state{CA}
  \postcode{95053}
  \country{USA}
}
\email{tebesu@scu.edu}

\author{Bin Shen}
\affiliation{%
  \institution{Google}
  \streetaddress{111 8th Avenue}
  \city{New York} 
  \state{NY} 
  \postcode{10011}
  \country{USA}
}
\email{bshen@google.com}

\author{Yi Fang}
\affiliation{%
  \institution{Santa Clara University}
  \institution{Department of Computer Engineering}
  \city{Santa Clara}
  \state{CA}
  \postcode{95053}
  \country{USA}
}
\email{yfang@scu.edu}

\begin{abstract}
Recommendation systems play a vital role to keep users engaged with personalized content in modern online platforms. Deep learning has revolutionized many research fields and there is a recent surge of interest in applying it to collaborative filtering (CF). However, existing methods compose deep learning architectures with the latent factor model ignoring a major class of CF models, neighborhood or memory-based approaches. We propose Collaborative Memory Networks (CMN), a deep architecture to unify the two classes of CF models capitalizing on the strengths of the global structure of latent factor model and local neighborhood-based structure in a nonlinear fashion. Motivated by the success of Memory Networks, we fuse a memory component and neural attention mechanism as the neighborhood component. The associative addressing scheme with the user and item memories in the memory module encodes complex user-item relations coupled with the neural attention mechanism to learn a user-item specific neighborhood. Finally, the output module jointly exploits the neighborhood with the user and item memories to produce the ranking score. Stacking multiple memory modules together yield deeper architectures capturing increasingly complex user-item relations. Furthermore, we show strong connections between CMN components, memory networks and the three classes of CF models. Comprehensive experimental results demonstrate the effectiveness of CMN on three public datasets outperforming competitive baselines. Qualitative visualization of the attention weights provide insight into the model's recommendation process and suggest the presence of higher order interactions.
\end{abstract}

%



\maketitle

\renewcommand{\citep}[1]{\citeauthor{#1} \cite{#1} }

\section{Introduction}
Recommendation systems are vital to keeping users engaged and satisfied with personalized recommendations in the age of information explosion. Users expect personalized content in modern E-commerce, entertainment and social media platforms but the effectiveness of recommendations are restricted by existing user-item interactions and model capacity. 
The ability to leverage higher order reasoning may help alleviate the problem of sparsity. A popular and successful technique, collaborative filtering (CF), establishes the relevance between users and items from past interactions (e.g., clicks, ratings, purchases) by assuming similar users will consume similar items.

CF can generally be grouped in three categories: memory or neighborhood-based approaches, latent factor models and hybrid models \cite{Ricci2011IntroRecSysHandbook, koren2008factorization}. Memory or neighborhood-based methods form recommendations by identifying groups or neighborhoods of similar users or items based on the previous interaction history. 
The simplicity of these models such as item $K$ nearest neighbor (KNN) have shown success in production systems at Amazon \cite{Linden2003AmazonCF}. Latent factor models such as matrix factorization project each user and item into a common low dimensional space capturing latent relations. Neighborhood methods capture local structure but typically ignore the mass majority of ratings available due to selecting at most $K$ observations from the intersection of feedback between two users or items \cite{koren2008factorization}. On the other hand, latent factor models capture the overall global structure of the user and item relationships but often ignore the presence of a few strong associations. The following weaknesses between the local neighborhood-based and global latent factor models lead to the development of hybrid models such as SVD++ \cite{koren2008factorization} and generalizations such as Factorization Machines \cite{rendle2010factorization} which integrate both neighborhood-based approaches and latent factor models to enrich predictive capabilities.

Recently, deep learning has made massive strides in many research areas obtaining state of the art performance in computer vision \cite{he2015delving}, question answering \cite{Weston2014MemoryN, Sukhbaatar2015EndToEndMN, Kumar2016AskMA, xiong2016dynamic}, learning programs \cite{graves2016DiffNeuralComp}, machine translation \cite{bahdanau2015-nmt-attn} and many other domains.
The successful integration of deep learning methods in recommendation systems have demonstrated the noticeable advantages of complex nonlinear transformations over traditional linear models \cite{zhang2017deep}. However, existing composite architectures incorporate the latent factor model ignoring the integration of neighborhood-based approaches in a nonlinear fashion.
Hence, we propose to represent the neighborhood-based component with a Memory Network \cite{Weston2014MemoryN, Sukhbaatar2015EndToEndMN} to capture higher order complex relations between users and items. An external memory permits encoding rich feature representations while the neural attention mechanism infers the user specific contribution from the community. 

We propose a unified hybrid model which capitalizes on the recent advances in Memory Networks and neural attention mechanisms for CF with implicit feedback. The memory component allows read and write operations to encode complex user and item relations in the internal memory. An associative addressing scheme acts as a nearest neighborhood model finding semantically similar users based on an adaptive user-item state. The neural attention mechanism places higher weights on specific subsets of users who share similar preferences forming a collective neighborhood summary. %
Finally, a nonlinear interaction between the local neighborhood summary and the global latent factors\footnote{We use the terms user/item latent factors, memories and embeddings interchangeably.} derives the ranking score. Stacking multiple memory components allows the model to reason and infer more precise neighborhoods further improving performance.

Our primary contributions can be summarized as follows:
\begin{itemize}
    \item We propose Collaborative Memory Network (CMN) inspired by the success of memory networks to address implicit collaborative filtering. CMN is augmented with an external memory and neural attention mechanism. The associative addressing scheme of the memory module acts as a nearest neighborhood model identifying similar users. The attention mechanism learns an adaptive nonlinear weighting of the user's neighborhood based on the specific user and item. The output module exploits nonlinear interactions between the adaptive neighborhood state jointly with the user and item memories to derive the recommendation.
    
    \item We reveal the connection between CMN and the two important classes of collaborative filtering models: the latent factor model and neighborhood-based similarity model. Furthermore, we reveal the advantages of the nonlinear integration fusing the two types of models yielding a hybrid model.
    
    \item Comprehensive experiments on three public datasets demonstrate the effectiveness of CMN against seven competitive baselines. Multiple experimental configurations confirm the added benefits of the memory module 
    \footnote{Source code available at:  \url{http://github.com/tebesu/CollaborativeMemoryNetwork}}.

    \item Qualitative visualizations of the attention weights provide insight into the memory component providing supporting evidence for deeper architectures to capture higher order complex interactions.
\end{itemize}

\section{Related Work}\label{related}

\subsection{Deep Learning in Recommendation Systems}
Recently, a surge of interest in applying deep learning to recommendation systems has emerged. Among the early works, \citep{salakhutdinov2007restricted} address collaborative filtering by applying a two layer Restricted Boltzmann Machine modeling tabular movie ratings. Autoencoders have been a popular choice of deep learning architecture for recommender systems \cite{Wu2016CDAE, wang2015collaborative, Sedhain2015, Li2015DCF, Zhang2017AutoSvd}. The autoencoder acts as a nonlinear decomposition of the rating matrix replacing the traditional linear inner product. For example, AutoRec \cite{Sedhain2015} decomposes the rating matrix with an autoencoder followed by reconstruction to directly predict ratings obtaining competitive results on numerous benchmark datasets. Collaborative denoising autoencoders (CDAE) \cite{Wu2016CDAE} address top-$n$ recommendation by integrating a user-specific bias into an autoencoder demonstrating CDAE can be seen as a generalization of many existing collaborative filtering methods. \citep{Li2015DCF} adopt a marginalized denoising autoencoder to diminish the computational costs associated with deep learning. Employing two autoencoders, one for item content and the other for user content bridged with user and item latent factors. AutoSVD++ \cite{Zhang2017AutoSvd} extends the original SVD++ model with a contrastive autoencoder to capture auxiliary item information. A hierarchical Bayesian model \cite{wang2015collaborative} bridges matrix factorization with the deepest layer of a stacked denoising autoencoder leveraging item content in the process.

Neural Matrix Factorization \cite{NeuralCF-2017} address implicit feedback by jointly learning a matrix factorization and a feedforward neural network. The outputs are then concatenated before the final output to produce an interaction between the latent factors and the nonlinear factors. \citep{Ebesu2017} address the item cold-start problem by tightly coupling a deep neural network (DNN) for item content and pairwise matrix factorization for rating decomposition. The DNN constructs a robust representation of the item content and item latent factor for new items. 
\citep{cheng2016wide} jointly train a logistic regression and a DNN to leverage the generalization aspects of deep learning and specificity of generalized linear models for mobile app recommendations in the Google Play store.

Convolutional neural networks (CNN) in recommendation systems have been used to capture localized item feature representations of music \cite{Oord2013DeepCM}, text \cite{Seo2017ICN, Kim2016ConvMF} and images \cite{zhang2017deep}. Previous methods represent text as bag-of-words representations, CNN overcomes this limitation by learning weight filters to identify the most prominent phrases within the text. The sequential nature of recurrent neural networks (RNNs) provides desirable properties for time-aware \cite{Wu2017RecurrentRecNet} and session-based recommendation systems \cite{hidasi2015session}. For example, Recurrent Recommender Networks \cite{Wu2017RecurrentRecNet} capture temporal aspects with a user and item Long Short Term Memory (LSTM) cell coupled with stationary factors to identify movie popularity fluctuations.
\citep{Jannach2017SessionPlusNeighborhood} interpolate KNN with a session-based RNN \cite{hidasi2015session} demonstrating further performance gains. However, the interpolation scheme is a fixed weighting hyperparameter and lacks a nonlinear interaction to capture more complex relations. 
\citep{wang2017irgan} unify the generative and discriminative methodologies under the generative adversarial network \cite{GoodfellowDLBook} framework for web search, item recommendation, and question answering.

Attention mechanisms have been recently explored in recommender systems. \citep{gong2016hashtag} perform hashtag recommendation with a CNN augmented with an attention channel to concentrate on the most informative (trigger) words. However, a hyperparameter must be carefully set to control the threshold of triggering the word to be informative. \citep{huang2016hashtagMemoryNet} tackle the same task with an End-to-End Memory Network \cite{Sukhbaatar2015EndToEndMN} integrating a hierarchical attention mechanism over the user's previous tweets on a word and sentence level. \citep{chen2017attentive} incorporate multimedia content with an item level attention representing the user preferences and a component level attention to isolate item specific visual features. Similarly, \citep{Seo2017ICN} introduce a local and global attention mechanism over convolutions to model review text. \citep{xiao2017attentional} extend Factorization Machines \cite{rendle2010factorization} with an attention mechanism to learn the importance of each pairwise interaction rather than treating them uniformly. Most existing neural attention based methods rely on additional content or context information while our task is to study collaborative filtering. To the best of our knowledge, no prior work has employed the memory network architecture to address implicit feedback in the collaborative filtering setting.

\begin{figure*}[th]
\centering
    \includegraphics[width=6.75in]{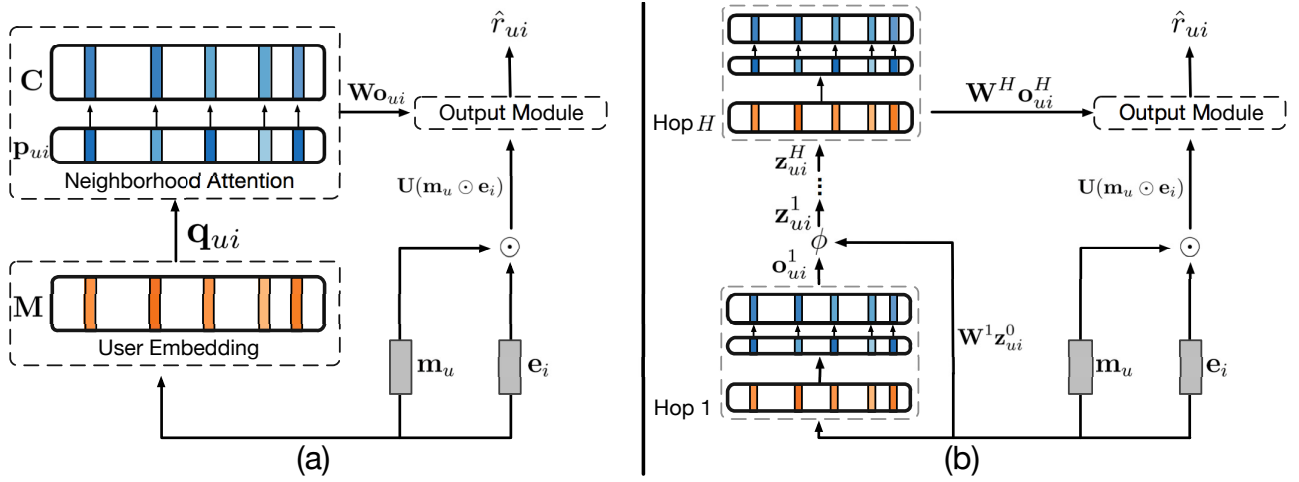}
    \caption{Proposed architecture of Collaborative Memory Network (CMN) with a single hop (a) and with multiple hops (b).}
\label{fig:arch}
\end{figure*}

\subsection{Memory Augmented Neural Networks}
We first provide a brief overview of the inner workings of memory-based architectures. Memory augmented neural networks, generally consist of two components: an external memory typically a matrix and a controller which perform operations on the memory (e.g., read, write, and erase). The memory component increases model capacity independent of the controller (typically a neural network) while providing an internal representation of knowledge to track long-term dependencies and perform reasoning. The controller manipulates these memories with either content-based or location-based addressing. 
Content-based or associative addressing finds a scoring function between the given question (query) and a passage of text, typically the inner product followed by the softmax operation leading to softly reading each memory location \cite{Weston2014MemoryN, Sukhbaatar2015EndToEndMN, Kumar2016AskMA, xiong2016dynamic, bahdanau2015-nmt-attn}. Performing a soft read over the memory locations allows the model to maintain
differentiation hence can be trained via backpropagation. The latter type of addressing (usually combined with content-based) performs sequential reads or random access \cite{graves2016DiffNeuralComp}.

The initial framework proposed by \citep{Weston2014MemoryN} demonstrated promising results to track long-term dependencies and perform reasoning over synthetic question answering tasks. \citep{Sukhbaatar2015EndToEndMN} alleviated the strong levels of supervision required to train the original memory network becoming an End-to-End system. The notion of attention is biologically motivated how humans do not uniformly process all information in a given task but focus on specific subsets of information. Attention mechanisms also provide a level of insight into the deep learning black box by visualizing the attention weights \cite{bahdanau2015-nmt-attn}. \citep{Kumar2016AskMA} improve upon the existing architecture by introducing an episodic memory component allowing for multiple passes or consultations of the memory before producing the final answer. The flexibility of the memory network architecture allows it to perform visual question answering \cite{xiong2016dynamic} and joint task learning for identifying the sentiment and the relation to target entity \cite{Li2017MemNetForAttitudeId}.

\section{Collaborative Memory Network}
In this section, we introduce our proposed model Collaborative Memory Network (CMN), see Figure \ref{fig:arch}a for a visual depiction of the architecture. At a high level, CMN maintains three memory states: an internal user-specific memory, an item-specific memory, and a collective neighborhood state. The architecture allows for the joint nonlinear interaction of the specialized local structure of neighborhood-based methods and the global structure of latent factor models. The associative addressing scheme acts as a nearest neighbor similarity function that learns to select semantically similar users based on the current item. The neural attention mechanism permits learning an adaptive nonlinear weighting function for the neighbor model, where the most similar users contribute higher weights at the output module.
We later extend the model to a deeper architecture by stacking multiple hops in Section \ref{sec:multi-hops} depicted in Figure \ref{fig:arch}b.

\subsection{User Embedding}
The memory component consists of a user memory matrix $\mathbf{M} \in \mathbb{R}^{P \times d}$ and an item memory matrix $\mathbf{E} \in \mathbb{R}^{Q \times d}$, where $P$ and $Q$ represents the number of users and items respectively and $d$ denotes the size (dimensionality) of each memory cell. 
Each user $u$ is embedded in a memory slot $\mathbf{m}_u \in \mathbf{M}$ storing her specific preferences. Similarly, each item $i$ corresponds to another memory slot $\mathbf{e}_i \in \mathbf{E}$ encoding the item's specific attributes. We form a user preference vector $\mathbf{q}_{ui}$ where each dimension $q_{uiv}$ is the similarity of the target user $u$'s level of agreement with user $v$ in the neighborhood given item $i$ as:
\begin{equation}\label{eqn:nn}
    q_{uiv} = \mathbf{m}_u^\mathsf{T} \mathbf{m}_v + \mathbf{e}_i^\mathsf{T} \mathbf{m}_v \quad \forall \; v \in N(i)
\end{equation}
where $N(i)$ represents the set of all users (neighborhood) who have provided implicit feedback for item $i$. 
We would like to point out $N(i)$ could be replaced or combined with $R(i)$ to handle the case of explicit feedback where $R(i)$ denotes the set of all users who provided explicit feedback for item $i$. 
The intuition is as follows, the first term computes compatibility between the target user and the users who have rated item $i$. The second term introduces the level of confidence user $v$ supports the recommendation of item $i$. Hence, the associative addressing scheme identifies the internal memories with the highest similarity of the target user $u$ with respect to neighborhood of users given the specific item.

\subsection{Neighborhood Attention}
The neural attention mechanism learns an adaptive weighting function to focus on a subset of influential users within the neighborhood to derive the ranking score. Traditional neighborhood methods predefine a heuristic weighting function such as Pearson correlation or cosine similarity and require specifying the number of users to consider \cite{Ricci2011IntroRecSysHandbook}. While factorizing the neighborhood partially alleviates this problem, it is still linear in nature \cite{kabbur2013fism}. Instead by learning a weighting function over the entire neighborhood, we no longer need to empirically predefine the weighting function or number of neighbors to consider. Formally, we compute the attention weights for a given user to infer the importance of each user's unique contribution to the neighborhood: 
\begin{equation}\label{eqn:attn}
    p_{uiv} = \frac{\exp(q_{uiv})}{\sum_{k \in N(i)} \exp(q_{uik})} \quad \forall \; v \in N(i)
\end{equation}
which produces a distribution over the neighborhood. The attention mechanism allows the model to focus on or place higher weights on specific users in the neighborhood while placing less importance on user's who may be less similar. Next we construct the final neighborhood representation by interpolating the external neighborhood memory with the attention weights:

\begin{equation}\label{eqn:weightedNeighborhood}
    \mathbf{o}_{ui} = \sum\limits_{v \in N(i)} p_{uiv} \mathbf{c}_v
\end{equation}
where $\mathbf{c}_v$ is another embedding vector for user $v$ which is called external memory in the original memory network framework \cite{Weston2014MemoryN}. Denoting the $v^{th}$ column of the embedding matrix $\mathbf{C}$ with the same dimensions as $\mathbf{M}$. 
The external memory allows the storage of long-term information pertaining specifically to each user's role in the neighborhood. 
In other words, the associative addressing scheme identifies similar users within the neighborhood acting as a key to weight the relevant values stored in the memory matrix $\mathbf{C}$ via the attention mechanism.
The attention mechanism selectively weights the neighbors according to the specific user and item.
The final output $\mathbf{o}_{ui}$ represents a weighted sum over the neighborhood composed of the relations between the specific user, item and the neighborhood.

CMN captures the similarity of users and dynamically assigns the degrees of contribution to the collective neighborhood based on the target item rather than a predefined number of neighbors which may restrict generalization capacity. 
Furthermore, the attention mechanism reduces the bottleneck of encoding all information into each individual memory slot and allows the joint exploitation of the user and item observations.

\subsection{Output Module}
As noted earlier neighborhood models capture the local structure from the rating matrix via the neighbors while latent factor models identify the global structure of the rating matrix \cite{koren2008factorization}. Hence we consider the collective neighborhood state to capture localized user-item relations and the user and item memories to capture the global user-item interactions.
The output module smoothly integrates a nonlinear interaction between the local collective neighborhood state and the global user and item memories. 
Existing models lack the nonlinear interaction between the two terms potentially limiting the extent of captured relations \cite{Zhang2017AutoSvd, chen2017attentive}. For a given user $u$ and item $i$ the ranking score is given as:

\begin{equation}\label{eqn:predict}
    \hat r_{ui} = \mathbf{v}^\mathsf{T} \phi \big(\mathbf{U}(\mathbf{m}_u \odot \mathbf{e}_i) + \mathbf{W}\mathbf{o}_{ui} + \mathbf{b}\big)
\end{equation}
where $\odot$ is the elementwise product; $\mathbf{W, U} \in \mathbb{R}^{d \times d}$; and $\mathbf{v, b} \in \mathbb{R}^d$ are parameters to be learned. 
We first apply the elementwise product between the user and item memories followed by a linear projection with $\mathbf{U}$, subsequently introducing a skip-connection thus reducing the longest path from the output to input. Skip-connections have been shown to encourage the flow of information and ease the learning process \cite{he2016resnet}. In this way, the model can better correlate the specific target addresses (user and item memories) with the ranking score to propagate the appropriate error signals.
We further motivate this choice by demonstrating its connection to the latent factor model (Section \ref{sec:lfm}). Similarly, the final neighborhood representation $\mathbf{o}_{ui}$ is projected to a latent space with $\mathbf{W}$ then combined with the previous term followed by a nonlinear activation function $\phi(\cdot)$.
Empirically we found the rectified linear unit (ReLU) $\phi(x) = \max(0, x)$ to work best due to its nonsaturating nature and suitability for sparse data \cite{GoodfellowDLBook, he2016resnet}.

Our proposed model provides the following advantages. First, consider the case where the amount of feedback for a given user is sparse, we can leverage all users who have rated the item to gain additional insight about the existing user and item relations. Second, the neural attention mechanism adjusts the confidence of each user's contribution to the final ranking score dependent on the specific item. Finally, the nonlinear interaction between the local neighborhood and global latent factors provide a holistic view of the user-item interactions.

\subsection{Multiple Hops}\label{sec:multi-hops}
We now extend our model to handle an arbitrary number of memory layers or hops. Figure \ref{fig:arch}b (right) illustrates CMN's architecture with multiple hops. Each hop queries the internal user memory and item memory followed by the attention mechanism to derive the next collective neighborhood state vector.
The first hop may introduce the need to acquire additional information. Starting from the second hop, the model begins to take into consideration the collective user neighborhood guiding the search for the representation of the community preferences. Each additional hop repeats this step considering the previous hop's newly acquired information before producing the final neighborhood state.
In other words, the model has the chance to look back and reconsider the most similar users to infer more precise neighborhoods. More specifically, multiple memory modules are stacked together by taking the output from the $h^{th}$ hop as input to the $(h+1)^{th}$ hop.
Similar to \cite{Sukhbaatar2015EndToEndMN, Li2017MemNetForAttitudeId,xiong2016dynamic} we apply a nonlinear projection between hops:

\begin{equation}
\mathbf{z}^{h}_{ui} = \phi(\mathbf{W}^{h}\mathbf{z}^{h-1}_{ui} + \mathbf{o}^h_{ui} + \mathbf{b}^h) 
\end{equation}
where $\mathbf{W}^h$ is a square weight matrix mapping the user preference query $\mathbf{z}^{h-1}_{ui}$ to a latent space coupled with the existing information from the previous hop followed by a nonlinearity and the initial query $\mathbf{z}^{0}_{ui} = \mathbf{m}_u + \mathbf{e}_i$. Intuitively, the initial consultation of the memory may introduce the need for additional information to infer more precise neighborhoods. The nonlinear transformation updates the internal state then solicits the user neighborhood:

\begin{equation}
q_{uiv}^{h+1} = (\mathbf{z}^{h}_{ui})^\mathsf{T}\mathbf{m}_v \quad \forall \; v \in N(i)
\end{equation}
The newly formed user preference vector then recomputes the compatibility between the target user and the neighborhood followed by the adaptive attention mechanism producing an updated collective neighborhood summary. This process is repeated for each hop yielding an iterative refinement.
The output module receives the weighted neighborhood vector from the last ($H^{th}$) hop to produce the final recommendation.

\subsection{Parameter Estimation}
Since our objective is to study implicit feedback which is more pervasive in practice and can be collected automatically (e.g. clicks, likes). In the case of implicit feedback, the rating matrix contains a 1 if the item is observed and 0 otherwise.
We opt for the pairwise assumption, where a given user $u$ prefers the observed item $i^+$ over unobserved or negative item $i^-$. The traditional pointwise approach assumes the user is not interested in the item $i^-$ but in reality may not be aware of the item. We can form triplet preferences ($u, i^+, i^-$) since the number of preference triplets is quadratic in nature we uniformly sample a ratio of positive items to negative items which we further investigate in Section \ref{sec:neg_sampling}.
We leverage the Bayesian Personalized Ranking (BPR) optimization criterion \cite{Rendle2009} as our loss function which approximates AUC (area under the ROC curve):

\begin{equation}\label{eqn:loss}
\mathcal{L} = -\sum\limits_{(u,i^+,i^-)} \log \sigma(\hat r_{ui^+} - \hat r_{ui^-}) 
\end{equation}
where $\sigma(x) = 1 / \big(1+\exp(-x)\big)$ is the logistic sigmoid function. 
It is worth noting we are not restricted to setting $\sigma(x)$ as the logistic sigmoid function. Other pairwise probability functions such as the Probit function can be used as in \cite{Ebesu2017}.
Since the entire architecture is differentiable, CMN can be efficiently trained with the backpropagation algorithm.
To reduce the number of parameters we perform layerwise weight tying sharing all embedding matrices across hops \cite{Sukhbaatar2015EndToEndMN, Li2017MemNetForAttitudeId,xiong2016dynamic}.

\subsection{Computational Complexity}
The computational complexity for a forward pass through CMN for a user is $O\big(d|N(i)| + d^2 + d\big)$ where $|N(i)|$ denotes the size of the neighborhood for item $i$ and $d$ is the embedding size. The first term $O\big(d|N(i)|\big)$ is the cost for computing the user preference vector and the latter terms correspond to the final interactions in the output module.
Each additional hop introduces $O\big(d|N(i)| + d^2\big)$ complexity.
During training two forward passes are computed one for the observed positive item and the second for the negative unobserved item. Parameters can be updated via backpropagation with the same complexity. In real-world datasets, $|N(i)|$ is usually slightly larger than or comparable to $d$, and thus the primary computational complexity is computing $O(d|N(i)|)$. The cost is reasonable since other deep learning methods such as CDAE \cite{Wu2016CDAE} compute a forward pass in $O\big(Qd\big)$ where $Q$ is the total number of items. The proposed memory module only computes the similarity with the target user's neighbors (not over all users) and $|N(i)|$ is often less than or comparable to $Q$. Thus, the training in CMN is quite efficient. In practice, prefiltering techniques can be used to limit the number of neighbors to the top-$K$ because not all neighbors may be indicative in contributing to the final prediction \cite{Ricci2011IntroRecSysHandbook}. Since the purpose of our study is to understand the characteristics of CMN, we leave prefiltering techniques to future work.

Recommendation can be performed by computing the predicted ranking score (Eqn. \ref{eqn:predict}) for a given user and item with a single pass through the network. The item with the highest value is recommended to the user. The computational complexity for runtime recommendation is the same with that of the single forward pass during training.

\subsection{Relation to Existing Models}\label{sec:relation}
CMN consists of components which can be interpreted in terms of the three classes of collaborative filtering methods. We show the connection with the latent factor model and neighborhood-based similarity models, and finally the relation to hybrid models such as SVD++. We conclude the section by drawing parallels between memory networks and CMN.

\subsubsection{Latent Factor Model}\label{sec:lfm}
The latent factor model discovers hidden relations by decomposing the ratings matrix into two lower rank matrices. By omitting the neighborhood term and bias, and further setting $\mathbf{U}$ to the identity matrix Eqn. \ref{eqn:predict} becomes the following:
\begin{equation}\label{eqn:gmf}
\hat r_{ui} = \mathbf{v}^\mathsf{T} \phi(\mathbf{m}_u \odot \mathbf{e}_i)
\end{equation}
which leads to a generalized matrix factorization (GMF) model \cite{NeuralCF-2017}. Removing the nonlinearlity by setting $\phi(\cdot)$ to the identity function and constraining $\mathbf{v}$ to $\mathbbm{1}$ vector of all ones, we recover matrix factorization. Under our pairwise loss function (Eqn. \ref{eqn:loss}) we recover BPR \cite{Rendle2009}.

\subsubsection{Neighborhood-based Similarity Model}
The objective of neighborhood-based similarity models are to estimate a user-user\footnote{Equivalently switching users with items yields item-based methods.} similarity matrix $\mathbf{S} \in \mathbb{R}^{P \times P}$. For each user who rated item $i$ an aggregated similarity score produces the confidence of recommending the item. The general form of neighborhood similarity models are:
\begin{equation}\label{eqn:simModels}
    \hat r_{ui} = \alpha \sum\limits_{v \in N(i)} \mathbf{S}_{uv}
\end{equation}
where $\alpha$ is a normalization term to weight the ranking score. In the simplest case, the normalization term is set to $|N(i)|^{-\rho}$ where $\rho$ is a hyperparameter controlling the level of similarity required to obtain a high score. In KNN, the neighborhood $N(i)$ is restricted to be the weighted combination of the $K$ most similar users and the similarity matrix $\mathbf{S}$ is approximated with a heuristically predefined function such as Pearson correlation or cosine similarity. Another approach is to learn the similarity function by approximating $\mathbf{S}$ \cite{ning2011slim,kabbur2013fism}.
In our case, the attached memory module from Eqn. \ref{eqn:weightedNeighborhood} acts as the neighborhood similarity matrix. If we designate the attention mechanism as a predefined normalization term the user-user similarity matrix is then factorized as $\mathbf{S} = \mathbf{CC}^\mathsf{T}$.
Using the prediction rule from Eqn. \ref{eqn:simModels} the memory module yields a user-based variant of FISM and under the BPR loss function we recover FISMauc \cite{kabbur2013fism}.

\subsubsection{Hybrid Model}
We have shown the connection between the components of CMN and the two classes of collaborative filtering models. Hybrid models such as SVD++ \cite{koren2008factorization} contains two general terms, a user-item latent factor interaction and a neighborhood component. The output module (Eqn. \ref{eqn:predict}) smoothly integrates the latent factors and the similarity or neighborhood terms together leading to a hybrid model.
\begin{equation}
    \hat r_{ui} = \mathbf{v}^\mathsf{T} \phi\big(\overbrace{\mathbf{m}_u \odot \mathbf{e}_i}^{\text{Latent Factors}} + \underbrace{\sum_{v\in N(i)} p_{uiv}\mathbf{c}_v}_{\text{Neighborhood}}\big)
\end{equation}
We remove the projection matrices and bias terms for clarity. We can see the global interaction from the latent factors consist of the user and item memories. The memory module represents the localized neighborhood component and the neighborhood normalization term is replaced with the adaptive attention mechanism pushed inside the summation becoming a user-item specific weighting scheme.
Unlike SVD++, our hybrid model allows for complex nonlinear interactions between the two terms to model the diverse tastes of users.

\subsubsection{Memory Networks} Traditional memory networks address the task of question answering. A short story or passage of text is provided along with a question for which the answer can be derived by leveraging some form of reasoning. 
If we pose recommendation as a question answering problem we are asking how likely will this user enjoy the item where the user neighborhood is the story and the output ranking score is the answer.
Continuing our analogy, each word in the story acts as a user in the neighborhood providing supporting evidence for the recommendation.

\begin{table}[t]
\centering
\begin{tabular}{|lcccc|}
\hline
\bf{Dataset}  & \bf{Ratings}  & \bf{Users}  & \bf{Items}    & \bf{Sparsity} \\ \hline
\textit{Epinions} & 664,823 & 40,163 & 139,738  & 99.98\% \\ \hline
\textit{citeulike-a}& 204,987  & 5,551 & 16,980  & 99.78\% \\\hline
\textit{Pinterest} & 1,500,809 & 55,187 & 9,916  & 99.73\%\\\hline
\end{tabular}
\caption{Dataset statistics.}
\label{table:stats}
\end{table}

\begin{table*}[]
    \centering
    \small
    \begin{tabular}{|l|cccc|cccc|cccc|}\hline
    {} & \multicolumn{4}{c|}{\textit{Epinions}} & \multicolumn{4}{c|}{\textit{citeulike-a}} & \multicolumn{4}{c|}{\textit{Pinterest}} \\ \hline
{} &     HR@5 &     HR@10 &   NDCG@5 &   NDCG@10 &     HR@5 &     HR@10 &   NDCG@5 &   NDCG@10 & HR@5 &     HR@10 &   NDCG@5 &   NDCG@10 \\\hline
KNN & 0.1549 & 0.1555 &  0.1433 &   0.1435 & 0.6990 & 0.7348 &  0.5789 &   0.5909 & 0.5738 & 0.8376 &  0.3450 &   0.4310 \\
FISM   & 0.5542 & 0.6717 &  0.4192 &   0.4573 & 0.6727 & 0.8072 &  0.5106 &   0.5545 & 0.6783 & 0.8654 &  0.4658 &   0.5268 \\
BPR    & 0.5584 & 0.6659 &  0.4334 &   0.4683 & 0.6547 & 0.8083 &  0.4858 &   0.5357 & 0.6936 & 0.8674 &  0.4912 &   0.5479 \\
SVD++  & 0.5628 & 0.6754 &  0.4112 &   0.4477 & 0.6952 & 0.8199 &  0.5244 &   0.5649 & 0.6951 & 0.8684 &  0.4796 &   0.5362\\
GMF    & 0.5365 & 0.6562 &  0.4015 &   0.4404 & 0.7271 & 0.8326 &  0.5689 &   0.6034 & 0.6726 & 0.8505 &  0.4737 &   0.5316 \\
CDAE   & 0.5666 & 0.6844 &  0.4333 &   0.4715 & 0.6799 & 0.8103 &  0.5106 &   0.5532 & 0.7008 & 0.8722 &  0.4966 &   0.5525 \\
NeuMF    & 0.5500 & 0.6660 &  0.4214 &   0.4590 & 0.7629 & 0.8647 &  0.5985 &   0.6316 & 0.7041 & 0.8732 &  0.4978 &   0.5530 \\
CMN-1  & 0.5962 & 0.6943 &  0.4684 &   0.5003 & 0.6692 & 0.7809 &  0.5213 &   0.5575 & 0.6984 & 0.8662 &  0.4960 &   0.5507    \\
CMN-2  & 0.6017$\dagger$ & \textbf{0.7007}$\dagger$ &  0.4724$\dagger$ &   0.5045$\dagger$ & \textbf{0.7959}$\dagger$ & \textbf{0.8921}$\dagger$ &  0.6185$\dagger$ &   0.6500$\dagger$ & 0.7267$\dagger$ & 0.8904$\dagger$ &  \textbf{0.5180}$\dagger$ &   0.5714$\dagger$ \\
CMN-3  & \textbf{0.6020}$\dagger$ & 0.6985$\dagger$ &  \textbf{0.4748}$\dagger$ &   \textbf{0.5062}$\dagger$ & 0.7932$\dagger$ & 0.8901$\dagger$ &  \textbf{0.6234}$\dagger$ &   \textbf{0.6551}$\dagger$ & \textbf{0.7277}$\dagger$ & \textbf{0.8931}$\dagger$ &  0.5175$\dagger$ &   \textbf{0.5715}$\dagger$ \\
 \hline
    \end{tabular}
\caption{Experimental results for different methods on the \textit{Epinions}, \textit{citeulike-a} and \textit{Pinterest} datasets. Best results highlighted in bold. $\dagger$ indicates the improvement over baselines is statistically significant on a paired $t$-test ($p$ < 0.01).}
\label{table:baselines}
\end{table*}

\section{Experimental Results}

\subsection{Datasets}
We study the effectiveness of our proposed approach on three publicly available datasets.
The first dataset is collected from
\textit{Epinions}\footnote{\url{http://www.trustlet.org/downloaded_epinions.html}}
\cite{MassaEpinionsDataset} which provides an online service for users to share product feedback in the form of explicit ratings (1-5) and reviews. We convert the explicit ratings to implicit feedback as a 1 if the user has rated the item and 0 otherwise. 
The second dataset is \textit{citeulike-a}\footnote{\url{http://www.cs.cmu.edu/~chongw/data/citeulike/}} \cite{wang2011CTR} collected from CiteULike 
an online service which provides users with a digital catalog to save and share academic papers. User preferences are encoded as 1 if the user has saved the paper (item) in their library. The third dataset from \textit{Pinterest}\footnote{\url{http://sites.google.com/site/xueatalphabeta/}} \cite{geng2015learning} allows users to save or pin an image (item) to their board indicating a 1 or positive interaction otherwise a 0 and preprocessed according to \cite{NeuralCF-2017}.
Table \ref{table:stats} summarizes the statistics of the datasets.

\subsection{Evaluation}
We validate the performance of our proposed approach using the leave-one-out evaluation method following the prior work \cite{NeuralCF-2017,chen2017attentive,Rendle2009}. Closely following the setup from \citeauthor{NeuralCF-2017} \cite{NeuralCF-2017}, for each user we randomly hold out one item the user has interacted with and sample 100 unobserved or negative items to form the test set. The remaining positive examples form the training set.
If the user has only rated a single item we keep it in the training set to prevent the cold-start setting. We rank the positive item along with the 100 negative items and adopt two common ranking evaluation metrics \textit{Hit Ratio (HR)} and \textit{Normalized Discounted Cumulative Gain (NDCG)} \cite{Ricci2011IntroRecSysHandbook}. 
Intuitively, HR measures the presence of the positive item within the top $N$ and NDCG measures the items position in the ranked list and penalizes the score for ranking the item lower in the list.

\subsection{Baselines and Settings}
We compare our proposed approach against seven competitive baselines representing neighborhood-based; traditional latent factor models; hybrid model and deep learning-based models.

\begin{itemize}
    \item KNN \cite{Ricci2011IntroRecSysHandbook} is a neighborhood-based approach computing the cosine item-item similarity to provide recommendations.
    
    \item Factored Item Similarity Model (FISM)  \cite{kabbur2013fism} is a neighborhood-based approach factorizing the item-item similarity matrix into two low rank matrices optimizing the BPR loss function.
    
    \item Bayesian Personalized Ranking (BPR) \cite{Rendle2009} is a competitive pairwise matrix factorization for implicit feedback.
    
    \item SVD++ \cite{koren2008factorization} is a hybrid model combining the neighborhood-based similarity and the latent factor model.
    
    \item Generalized Matrix Factorization (GMF) \cite{NeuralCF-2017} is a nonlinear generalization of the latent factor model. We use the ReLU activation function and optimize the BPR loss function.

    \item Collaborative Denoising Auto Encoder (CDAE) \cite{Wu2016CDAE} is an item-based deep learning model for item ranking with a user specific bias.
    
    \item Neural Matrix Factorization (NeuMF) \cite{NeuralCF-2017} is a composite matrix factorization jointly coupled with a multilayer perceptron model for item ranking.
\end{itemize}

We would like to note that FISM \cite{kabbur2013fism} improves upon Sparse Linear Methods (SLIM) \cite{ning2011slim} by factorizing the item-item similarity matrix to handle missing entries hence we do not compare to SLIM. We exclude baselines utilizing additional information for fair comparison since our objective is to study implicit collaborative filtering without content or contextual information.

All hyperparameters are tuned according to the validation set. The validation set is formed by holding out one interaction per user from the training set \cite{NeuralCF-2017}.
We perform a grid search over each model's latent factors from $\{10, 20, 30, 40, 50\}$ and regularization terms $\{0.1, 0.01, 0.001\}$. In addition, we varied CDAE's corruption ratio from $\{0, 0.2, 0.4, 0.6, 0.8, 1.0\}$ and NeuMF's layers from $\{1, 2, 3\}$. The number of negative samples is set to 4.
Careful initialization of deep neural networks is crucial to avoid saddle points and poor local minima \cite{GoodfellowDLBook}. Thus, CMN initializes the user ($\mathbf{M}$) and item ($\mathbf{E}$) memory embeddings from a pretrained model according to Eqn. \ref{eqn:gmf}.
Remaining parameters are initialized according to \cite{he2015delving} which adapts the variance preserving Xavier initialization \cite{GoodfellowDLBook} for the ReLU activation function.
The gradient is clipped if the norm exceeds 5; $l_2$ weight decay of 0.1 with the exception of the user and item memories; and the mini-batch size is set to 128 for \textit{Epinions}, \textit{citeulike-a} and 256 for \textit{Pinterest}.
We adopt the RMSProp \cite{GoodfellowDLBook} optimizer with a learning rate of 0.001 and 0.9 for both decay and momentum. The default number of hops is set to 2 and memory or embedding size $d$ to 40, 50 and 50 for the \textit{Epinions}, \textit{citeulike-a} and \textit{Pinterest} datasets respectively. The effects of these hyperparameters are further explored in Sections \ref{sec:memory_size} and \ref{sec:neg_sampling}.

\begin{figure*}[!tbp]
    \centering
    \begin{subfigure}{.33\linewidth}
        \centering
        \includegraphics[width=\linewidth]{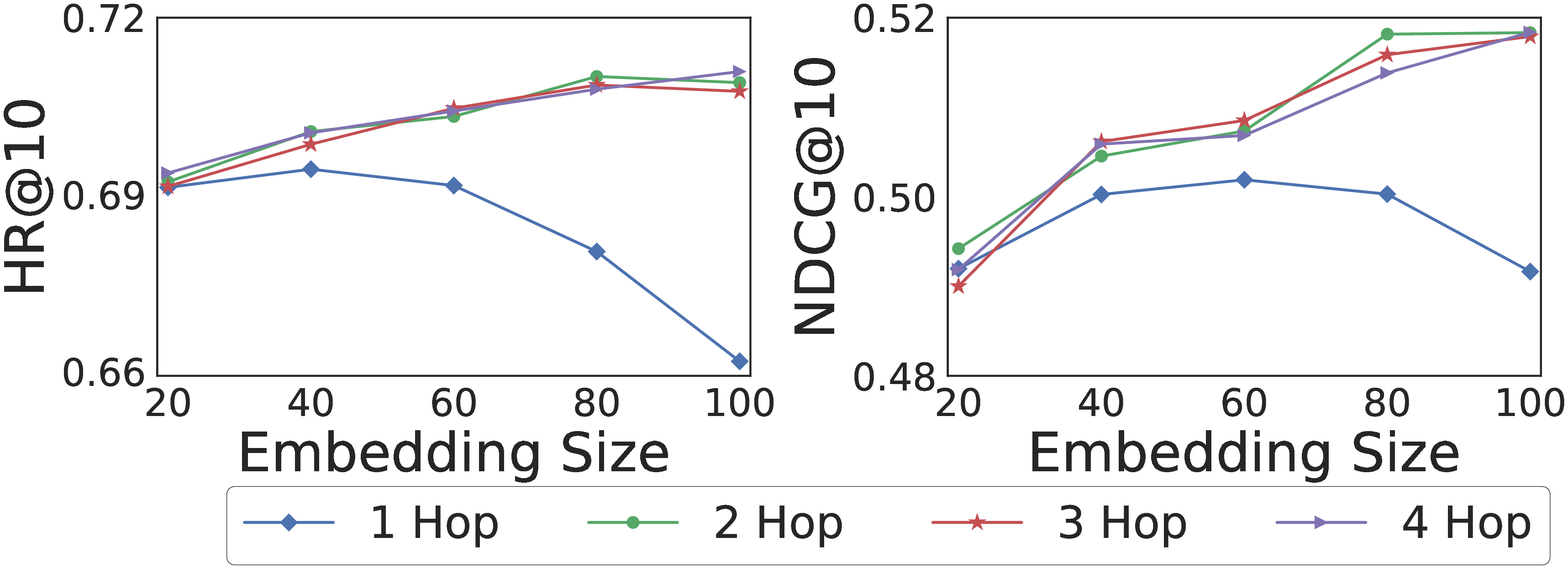}
        \caption{\textit{Epinions} dataset}
        \label{fig:epinions_embed}
    \end{subfigure}
    \begin{subfigure}{.33\linewidth}
        \centering
        \includegraphics[width=\linewidth]{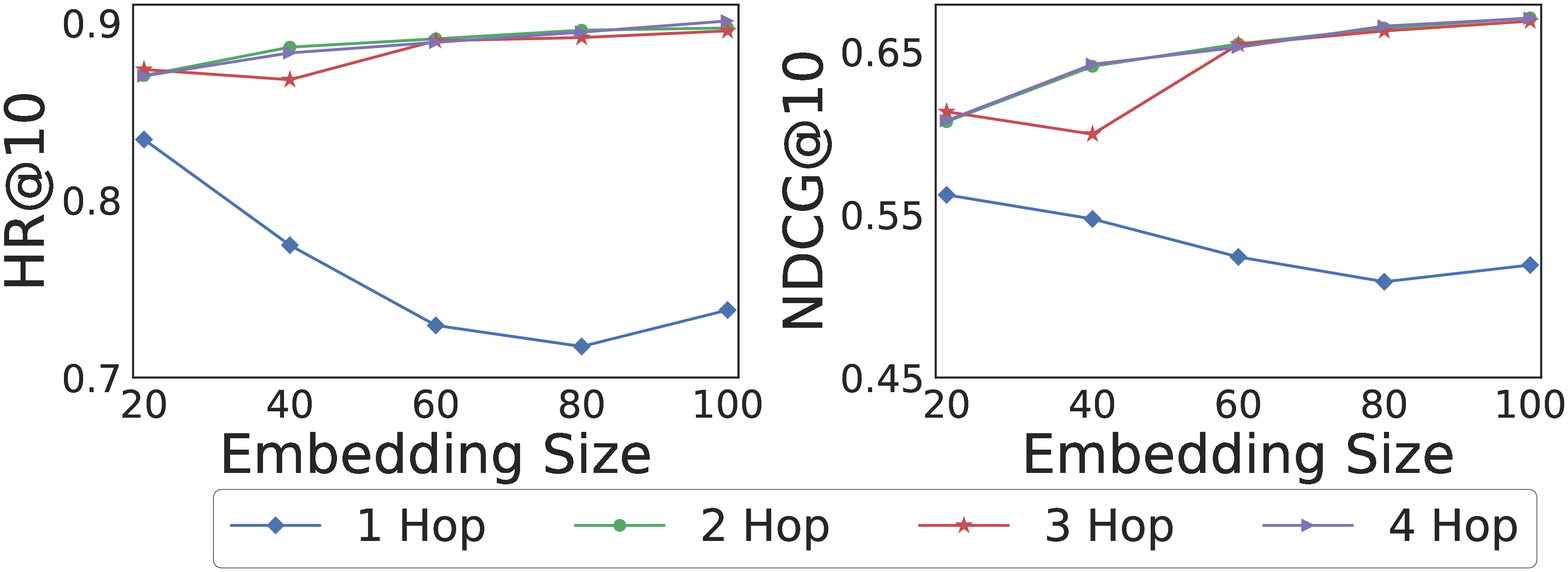}
        \caption{\textit{citeulike-a} dataset}
        \label{fig:citeulike_embed}
    \end{subfigure}
    \begin{subfigure}{.33\linewidth}
        \centering
        \includegraphics[width=\linewidth]{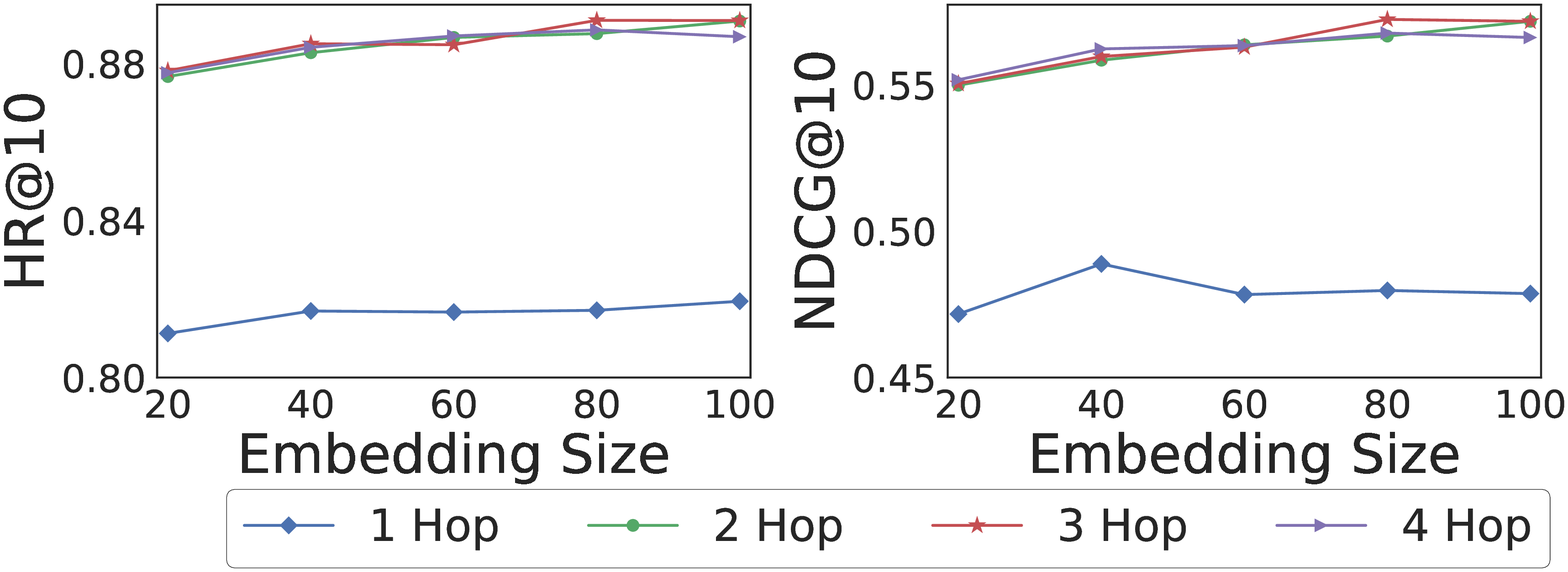}
        \caption{\textit{Pinterest} dataset}
        \label{fig:pinterest_embed}
    \end{subfigure}
  \caption{Experimental results for CMN varying the embedding size from 20-100 and hops from 1-4.}
\end{figure*}

\subsection{Baseline Comparison}
Table \ref{table:baselines} lists results of CMN with one, two and three hops along with the baselines for HR and NDCG with cut offs at 5 and 10 on the \textit{Epinions}, \textit{citeulike-a} and \textit{Pinterest} datasets. We denote CMN with one hop as `CMN-1', two hops with `CMN-2' and so on. At a high-level CMN variants obtain the best performance across both HR and NDCG at all cut offs for all datasets. 
We now provide a detailed breakdown of our results on the \textit{Epinions} dataset.
All baselines with the exception of KNN show competitive performance with each other across all metrics and cut offs.
Since CMN shares the same loss function with BPR, FISM and GMF, we can attribute the performance increase to the memory component. The application of a nonlinear transformation does not necessarily help as evident from BPR outperforming its nonlinear counterpart GMF.
KNN demonstrated the poorest performance particularly due to the restrictive ability to handle sparse data when only a few neighbors are present and a large number (139k) of items.
However, FISM's learned similarity function performs better since it can address missing entries in the item-item similarity matrix.
On the other hand, CMN can leverage the global structure of the latent factors encoded in the memory vectors and the additional memory component to infer complex user preferences. 
Furthermore, CMN's performance gains over CDAE, GMF and NeuMF portray the successful integration of the memory component and attention mechanism over existing nonlinear and deep learning-based methods.

\begin{table*}[t]
    \centering
    \small
    \begin{tabular}{|l|cccc|cccc|cccc|}\hline
    {} & \multicolumn{4}{c|}{\textit{Epinions}} & \multicolumn{4}{c|}{\textit{citeulike-a}} & \multicolumn{4}{c|}{\textit{Pinterest}} \\ \hline
{} &     \footnotesize{HR@5} &     \footnotesize{HR@10} &   \footnotesize{NDCG@5} &   \footnotesize{NDCG@10} &     \footnotesize{HR@5} &     \footnotesize{HR@10} &   \footnotesize{NDCG@5} &   \footnotesize{NDCG@10} & \footnotesize{HR@5} &     \footnotesize{HR@10} &   \footnotesize{NDCG@5} &   \footnotesize{NDCG@10} \\\hline
CMN & 0.6017 & 0.7007 &  0.4724 &   0.5045 & 0.7959 & 0.8921 &  0.6185 &   0.6500 & 0.7267 & 0.8904 &  0.5180 &   0.5714 \\
CMN-Attn & 0.5807 & 0.6948 &  0.4438 &   0.4809 & 0.7411 & 0.8589 &  0.5503 & 0.5887  & 0.6995 & 0.8773 &  0.4949 &    0.5530\\
CMN-Linear &  0.5954 & 0.6977 &   0.4660 &   0.4992 & 0.7721 & 0.8665 &  0.5974 &   0.6282 & 0.6992 & 0.8777 &  0.4951 &   0.5534 \\
CMN-Linear-Attn  & 0.5830 & 0.6937 &  0.4457 &   0.4816 & 0.7649 & 0.8676 &  0.5922 &   0.6256 & 0.6996 & 0.8775 &  0.4947 &   0.5527  \\
 \hline
    \end{tabular}
\caption{CMN variants without attention (CMN-Attn); linear activation with attention (CMN-Linear); and linear without attention (CMN-Linear-Attn).}
\label{table:arch}
\end{table*}

The denser \textit{citeulike-a} dataset contains fewer items than the \textit{Epinions} dataset leading to competitive performance from the item-based KNN method obtaining the strongest baseline NDCG@5. 
SVD++ outperforms the neighborhood-based FISM and latent factor BPR revealing the effectiveness of combining two approaches into a single hybrid model. The linear decomposition of the item-item similarity matrix in FISM may lack the expressiveness to capture complex preferences as suggested by the nonlinear GMF outperforming the linear BPR.
CMN demonstrates improved performance over the fixed neighborhood-based weighting of KNN and the learned linear similarity scheme of FISM indicating the additional nonlinear transformation and adaptive attention mechanism captures more complex semantic relations between users. 
CMN can be viewed as NeuMF by replacing the memory network component with a multilayer perceptron.
CMN outperforming NeuMF further establishes the advantage of the memory network component to identify complex interactions and iteratively update the internal neighborhood state. 

In the \textit{Pinterest} dataset, CMN with a single hop demonstrates competitive performance to baseline methods but with additional hops performance is further enhanced. SVD++ demonstrates competitive performance but may lack the full expressiveness of nonlinearity found in deep learning-based models to capture latent user-item relations. 
The larger dataset helps the two deep learning baselines to outperform the non-deep learning-based methods but the hybrid nature of CMN allows the joint nonlinear exploitation of the local neighborhood and global latent factors yielding additional performance gains.
Overall, CMN with two hops outperforms CMN with a single hop but supplementing additional hops greater than two did not provide significant advantages.

\subsection{Embedding Size}\label{sec:memory_size}
We illustrate the effect of varying the size of memory slots or embeddings and the number of hops for HR@10 and NDCG@10 on the \textit{Epinions} dataset in Figure \ref{fig:epinions_embed}. Since HR and NDCG show similar patterns, we focus our analysis on NDCG. The general trend shows a steady improvement as the embedding size increases with the exception of a single hop where an embedding size of 40 shows peak HR@10 performance followed by a degradation potentially due to overfitting. A single hop confines the model's expressiveness to infer and discriminate between the relevant and irrelevant information encoded in each memory cell. With a small embedding size of 20 increasing the number of hops provides negligible benefits but as the embedding size increases multiple hops show significant improvement over a single hop.

For the \textit{citeulike-a} dataset, Figure \ref{fig:citeulike_embed} portrays the best performance of a single hop at an embedding size of 20 followed by a degradation as model capacity increases which is somewhat similar to the \textit{Epinions} dataset. 
At three hops and an embedding size of 40 shows an unusual drop in performance potentially from finding a poor local minima due to the nonconvex nature of neural networks. 
Two and four hops show almost identical performance with at most a deviation of 0.3\% from each other on HR and NDCG.
In general, two hops demonstrates competitive performance against the three and four hop models across all embedding sizes.

The \textit{Pinterest} dataset in Figure \ref{fig:pinterest_embed} shows a similar trend of gradual performance gains as the embedding size increases but a single hop shows insufficient capacity to model complex user-item interactions. Unlike the results from the previous datasets the performance of a single hop does not degrade as the embedding size increases. The larger dataset may provide some implicit form of regularization to prevent overfitting.
Two, three and four hops show similar performance and incremental with larger embedding sizes. 
Identifying a sufficient number of hops initially takes precedence over the size of the embeddings. With a sufficient number of hops the embedding size can be increased yielding a trade off between computational cost and performance.
By introducing additional hops, CMN can better manipulate the memories and internal state to represent more complex nonlinear relations. 
Consistent with previous results, the addition of more than two hops do not show significant benefit.

\begin{figure*}[!tbp]
    \centering
    \begin{subfigure}{.33\linewidth}
        \centering
        \includegraphics[width=\linewidth]{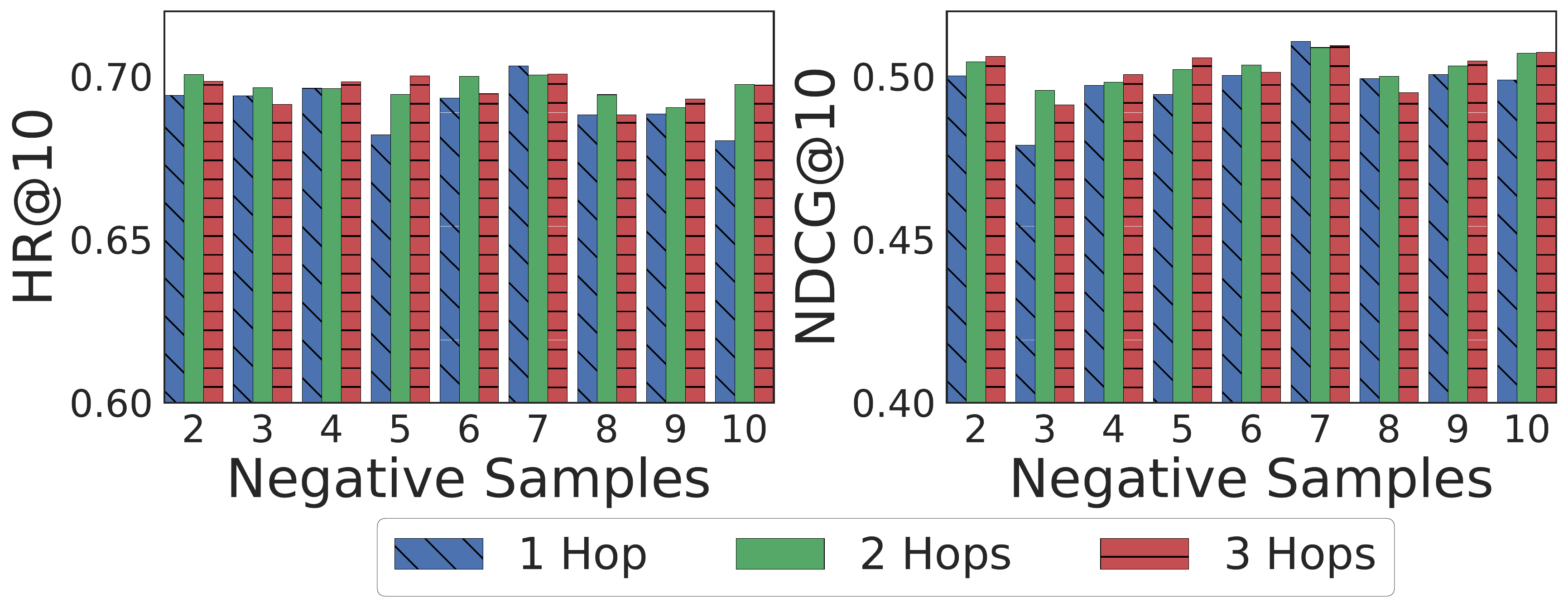}
        \caption{\textit{Epinions} dataset}
        \label{fig:epinions_neg}
    \end{subfigure}
    \begin{subfigure}{.33\linewidth}
        \centering
        \includegraphics[width=\linewidth]{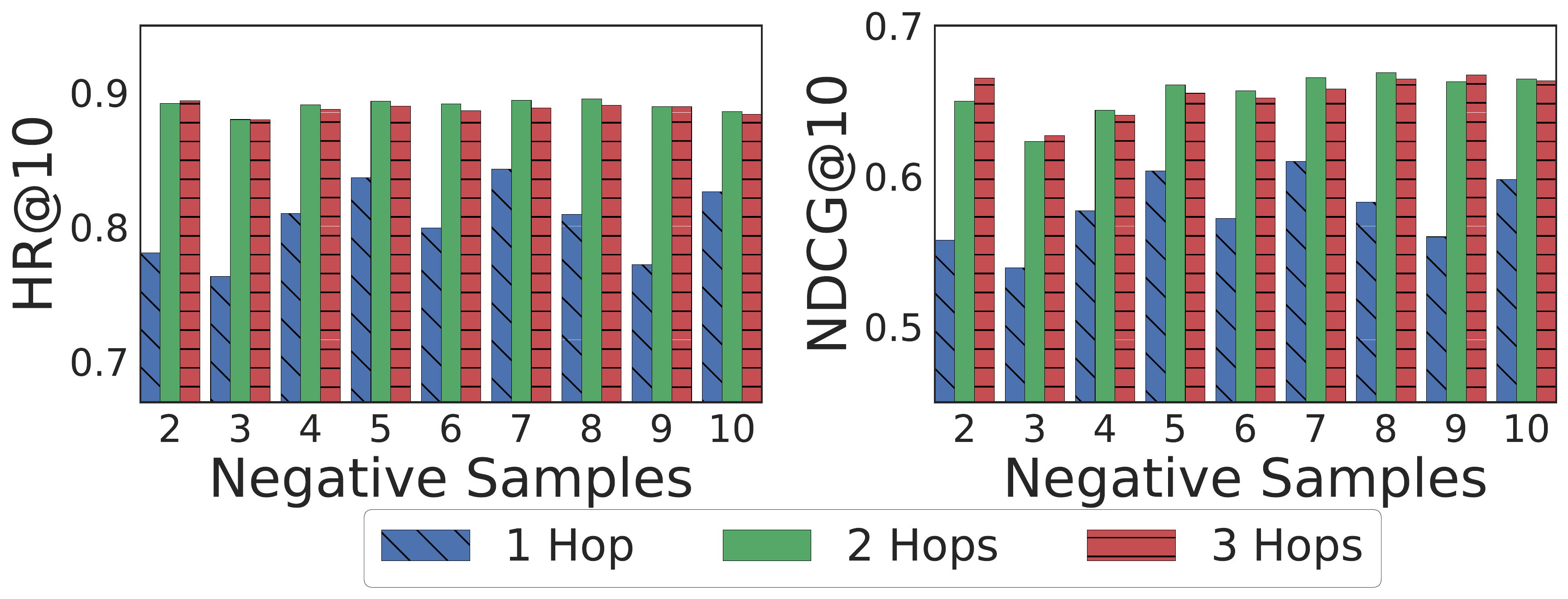}
        \caption{\textit{citeulike-a} dataset}
        \label{fig:citeulike_neg}
    \end{subfigure}
        \begin{subfigure}{.33\linewidth}
        \centering
        \includegraphics[width=\linewidth]{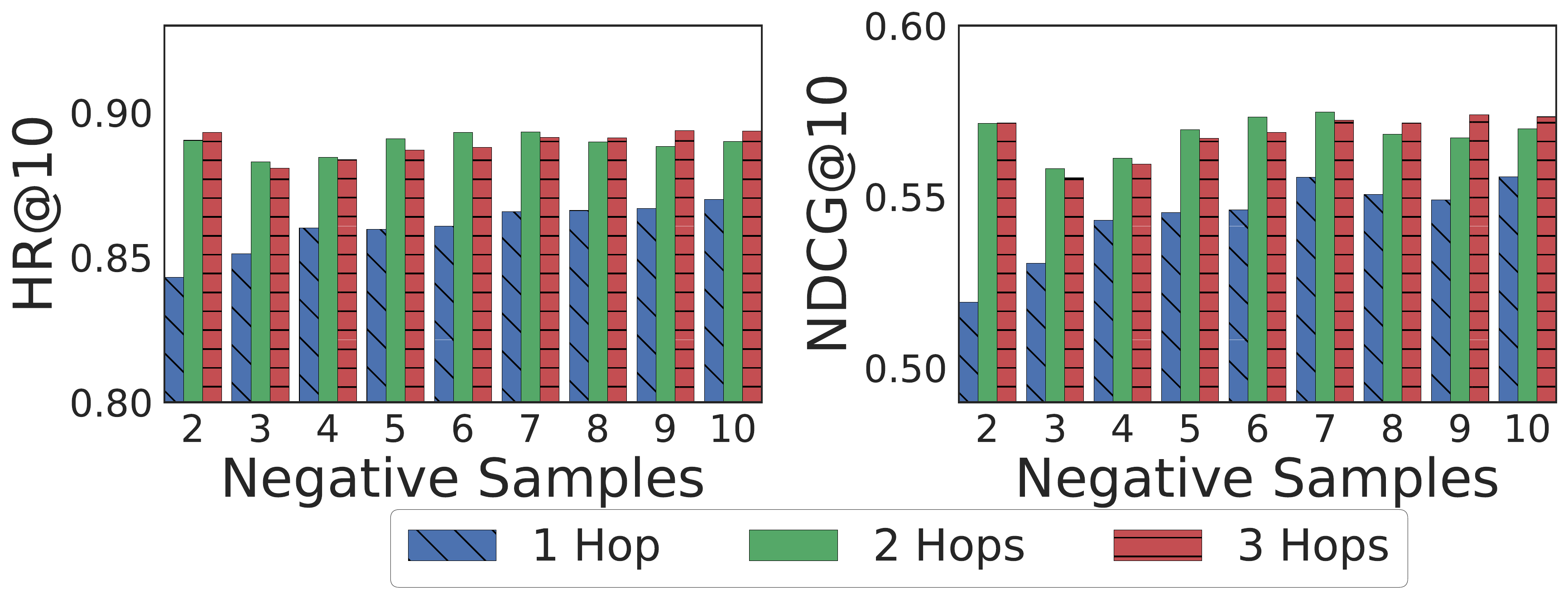}
        \caption{\textit{Pinterest} dataset}
        \label{fig:pinterest_neg}
    \end{subfigure}

  \caption{Experimental results for CMN varying the number of negative samples from 2-10 and hops from 1-3.}
\end{figure*}

\begin{figure*}[!tbp]
    \centering
    \begin{subfigure}{.33\linewidth}
        \centering
        \includegraphics[width=\linewidth]{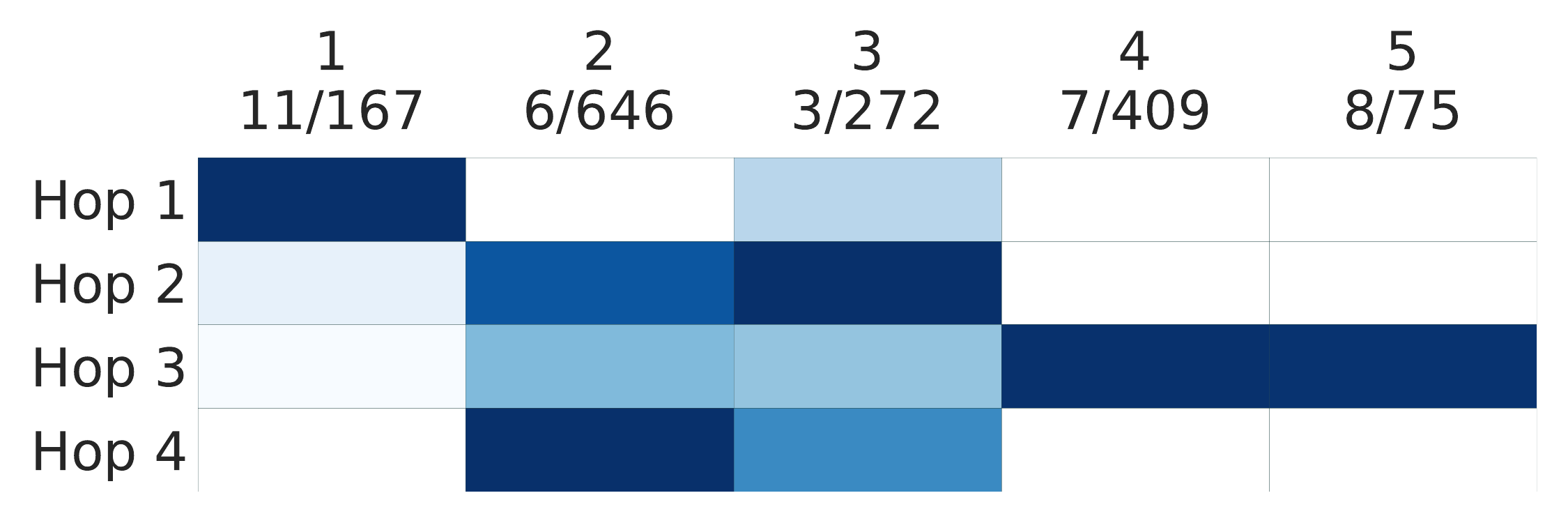}
        \caption{\textit{Epinions} dataset}
        \label{fig:epinions_attn}
    \end{subfigure}
    \begin{subfigure}{.33\linewidth}
        \centering
        \includegraphics[width=\linewidth]{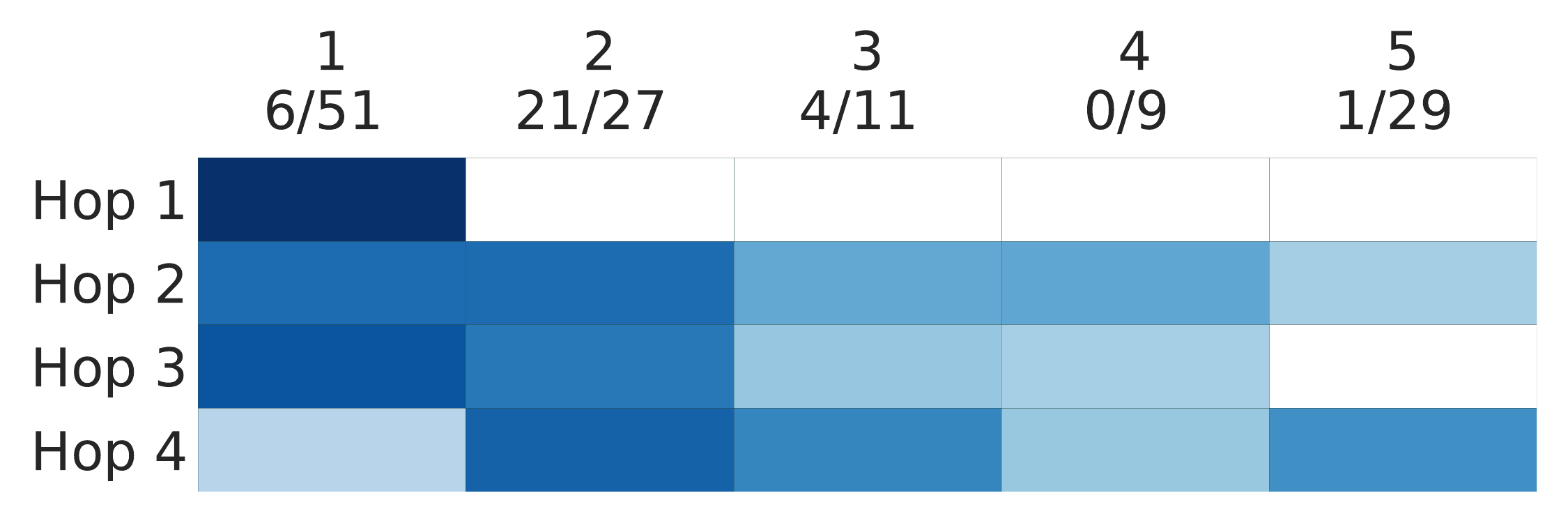}
        \caption{\textit{citeulike-a} dataset}
        \label{fig:citeulike_attn}
    \end{subfigure}
    \begin{subfigure}{.33\linewidth}
        \centering
        \includegraphics[width=\linewidth]{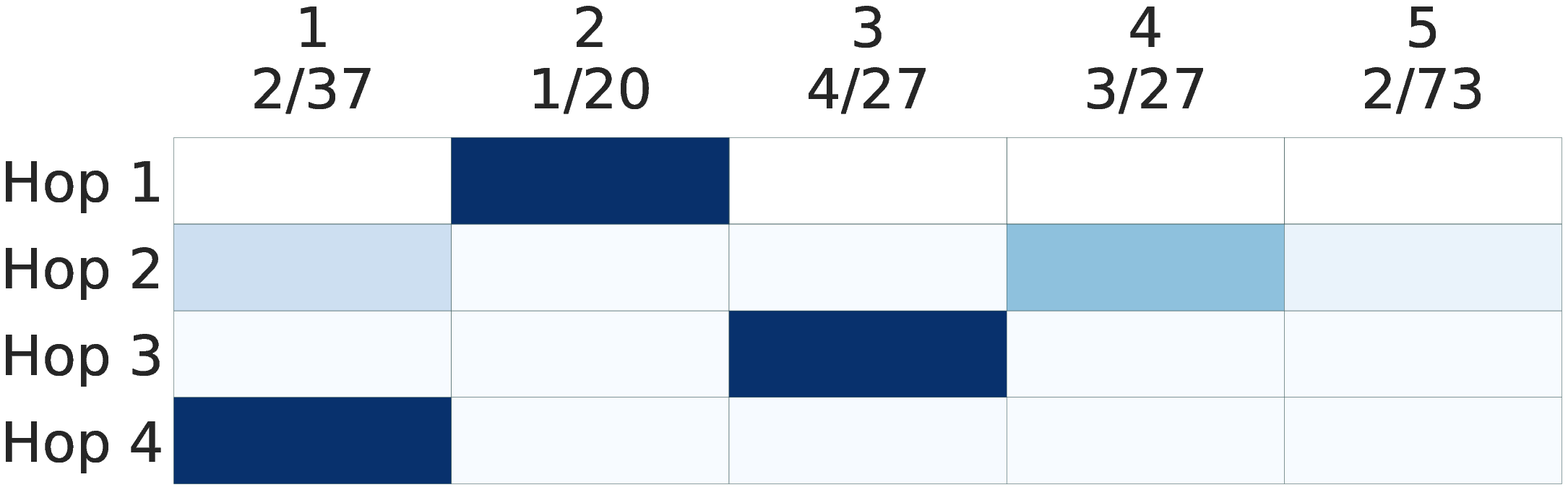}
        \caption{\textit{Pinterest} dataset}
        \label{fig:pinterest_attn}
    \end{subfigure}
      \caption{Heatmap of the attention weights over four hops. The color scale indicates the intensities of the weights, darker representing a higher weight and lighter a lower weight. Each column represents a user in the neighborhood labeled with the number of items in common with the target user / number of ratings in training set. For example, 11/167 indicates user 1 has 11 items corated with the current user $u$ and has rated a total of 167 items in the training set.}
    \label{fig:attn}
\end{figure*}

\subsection{Effects of Attention and Nonlinearity}
In this section, we seek to further understand the effect of individual components on performance. In Table \ref{table:arch}, the results for CMN without attention denoted `CMN-Attn' uniformly performs worse than with attention hinting at the effectiveness of the attention mechanism.
We also experimented with a linear version of CMN where all ReLU activation functions are set to the identity function denoted as `CMN-Linear'. The linear version with the attention mechanism generally outperforms the nonlinear version without attention. Further illustrating the effectiveness of the attention mechanism.
The final variation removes the attention mechanism from the linear version denoted as `CMN-Linear-Attn' which generally performs worse than the linear version with the attention mechanism. In general, removing the nonlinear transformation and attention mechanism in some variation yield similar performance on the \textit{Epinions} and \textit{Pinterest} datasets. In the \textit{citeulike-a} dataset the linear version with and without the attention mechanism show improvements over the nonlinear variant without the attention mechanism. This seems counter intuitive and may indicate a potential difficulty in finding a good local minima or a vanishing gradient problem which is consistent with the unusual drop in performance reported in Section \ref{sec:memory_size}.
CMN requires a combination from both the attention mechanism and nonlinear transformations to yield the best performance.

\subsection{Negative Sampling}\label{sec:neg_sampling}
In this section, we study the characteristics of varying the negative samples for CMN reporting HR@10 and NDCG@10. We exclude the results of four hops since the results were consistent with that of three hops. We also omit 1 negative sample since CMN was unable to distinguish between positive and negative samples leading to random performance.
Figure \ref{fig:epinions_neg} illustrates the performance of CMN varying the negative samples from 2-10 on the \textit{Epinions} dataset. A single hop shows fluctuations reporting low HR at 5 and 10 negative samples but outperforms the two and three hop versions with 7 negative samples. 
A single hop uniformly performs worse than the two and three hop counterparts in the \textit{citeulike-a} dataset presented in Figure \ref{fig:citeulike_neg}.
In Figure \ref{fig:pinterest_neg}, the results for a single hop on the \textit{Pinterest} dataset describes a general upward trend where performance  improves as the number of negative samples increase.
In both the \textit{citeulike-a} and \textit{Pinterest} datasets we observe two and three hops show comparable results and more stability to the number of negative samples while outperforming a single hop.
Overall, the performance of CMN is fairly stable with respect to the number of negative samples when at least two hops are present. 
Similar to the previous section on embedding size, we notice having at least two hops reduces the sensitivity to the hyperparameter.

\subsection{Attention Visualization}\label{sec:attn_vis}
Attention mechanisms allow us to visualize the weights placed on each user in the neighborhood with the hope of providing interpretable recommendations. We plot a heatmap of the weights from Eqn. \ref{eqn:attn} in Figure \ref{fig:attn}. The color scale represents the intensities of the attention weights, where a darker color indicates a higher value and lighter colors indicate a lower value. For ease of reference, we label each column representing a user in the neighborhood starting from 1 which may not necessarily reflect the true user id from the dataset. Furthermore, for each user we provide additional context in the form of user statistics. We denote 11/167 to indicate the user has rated a total of 167 items with 11 items observed in common with the target user in the training set. Since the size of the neighborhood can be large we limit the visualization to top 5 neighbors sorted by the highest aggregated attention values. We would like to point out that in some cases the attention weights can be small and hence not visually distinguishable.

The attention weights for a random user from the \textit{Epinions} dataset is portrayed in Figure \ref{fig:epinions_attn}. The user has a total of 49 neighbors thus we show only the top 5 neighbors due to space constraints. We can see all the top users attended to have at least a single item in common. The first hop places heavy levels of attention on user 1 and lightly on user 3. At two hops the attention on user 3 increases. Progressing to three hops the attention spread out across four users. 
Finally, at four hops user 2 and 3 have the highest weight with a balance of the number items in common with the target user and overall number of ratings observed suggesting these users may be the most influential in the recommendation process. 
As shown in previous sections performance generally increases with additional hops suggesting considering a combination of multiple users may be beneficial.

Figure \ref{fig:citeulike_attn} illustrates the attention weights over four hops for a random user from the \textit{citeulike-a} dataset with a total of 9 neighbors. We observe the first hop places a large amount of weight on a single user which may explain the poorer performance of CMN with a single hop. User 1 has the highest number of observations out of the neighborhood which may be a reasonable choice but it ignores other information that may be present from other users. Examining the weights of the second hop we see the attention is spread out across five users with a higher emphasis on users 1 and 2 who have the most items in common with the target user. Four out of the five users have a common item with the target user providing a strong indicator of the successful integration of the attention mechanism. Next, we focus on three hops which removes the attention over user 5 and reduces the intensities on user 4, 2 and 3. In the final hop, attention is returned to user 5 with stronger weights than in hop two. The overall attention levels shift around slightly but focus most heavily on user 2 which makes sense since it has the highest number of commonly rated items. Since user 4 has no items in common with the target user but large attention weights this warranted further investigation. We found user 4 to have at least one item in common with all other users in the neighborhood which may explain the attention placed on user 4 despite no corated items with the target user in the training data. This demonstrates the memory component captures higher level interactions within the neighborhood suggesting some form of transitive reasoning. 

Figure \ref{fig:pinterest_attn} illustrates the attention weights over four hops for a random user from the \textit{Pinterest} dataset. Similar to the previous visualization the first hop places heavy weights on a single user followed by a more dispersed weighting in the following hops. In hops two through four, a small amount of attention is placed upon each user which may not be visually distinguishable. Each hop allows CMN to examine the external memory and perhaps through some form of trial and error arrives at identifying the most useful neighbor as user 1 in hop four. 


\section{Conclusion}
We introduced a novel hybrid architecture unifying the strengths of the latent factor model and neighborhood-based methods inspired by Memory Networks to address collaborative filtering (CF) with implicit feedback. We reveal the connection between components of Collaborative Memory Network (CMN) the two important classes of CF models and draw parallels with the original memory network framework. Comprehensive experiments under multiple configurations demonstrate significant improvements over competitive baselines. Qualitative visualization of the attention weights provide insight into the model's recommendation process and suggest higher order transitive relations may be present. In future work, we hope to extend CMN to incorporate content and context information, tackle dialogue-based systems, 
and perform adversarial training \cite{GoodfellowDLBook}. 

\bibliographystyle{ACM-Reference-Format}
\bibliography{sigproc}


\begin{thebibliography}{41}


\ifx \showCODEN    \undefined \def \showCODEN     #1{\unskip}     \fi
\ifx \showDOI      \undefined \def \showDOI       #1{#1}\fi
\ifx \showISBNx    \undefined \def \showISBNx     #1{\unskip}     \fi
\ifx \showISBNxiii \undefined \def \showISBNxiii  #1{\unskip}     \fi
\ifx \showISSN     \undefined \def \showISSN      #1{\unskip}     \fi
\ifx \showLCCN     \undefined \def \showLCCN      #1{\unskip}     \fi
\ifx \shownote     \undefined \def \shownote      #1{#1}          \fi
\ifx \showarticletitle \undefined \def \showarticletitle #1{#1}   \fi
\ifx \showURL      \undefined \def \showURL       {\relax}        \fi
\providecommand\bibfield[2]{#2}
\providecommand\bibinfo[2]{#2}
\providecommand\natexlab[1]{#1}
\providecommand\showeprint[2][]{arXiv:#2}

\bibitem[\protect\citeauthoryear{Bahdanau, Cho, and Bengio}{Bahdanau
  et~al\mbox{.}}{2015}]%
        {bahdanau2015-nmt-attn}
\bibfield{author}{\bibinfo{person}{Dzmitry Bahdanau},
  \bibinfo{person}{Kyunghyun Cho}, {and} \bibinfo{person}{Yoshua Bengio}.}
  \bibinfo{year}{2015}\natexlab{}.
\newblock \showarticletitle{Neural machine translation by jointly learning to
  align and translate}. In \bibinfo{booktitle}{{\em ICLR}}.
\newblock


\bibitem[\protect\citeauthoryear{Chen, Zhang, He, Nie, Liu, and Chua}{Chen
  et~al\mbox{.}}{2017}]%
        {chen2017attentive}
\bibfield{author}{\bibinfo{person}{Jingyuan Chen}, \bibinfo{person}{Hanwang
  Zhang}, \bibinfo{person}{Xiangnan He}, \bibinfo{person}{Liqiang Nie},
  \bibinfo{person}{Wei Liu}, {and} \bibinfo{person}{Tat-Seng Chua}.}
  \bibinfo{year}{2017}\natexlab{}.
\newblock \showarticletitle{Attentive collaborative filtering: Multimedia
  recommendation with feature-and item-level attention}. In
  \bibinfo{booktitle}{{\em SIGIR}}.
\newblock


\bibitem[\protect\citeauthoryear{Cheng, Koc, Harmsen, Shaked, Chandra, Aradhye,
  Anderson, Corrado, Chai, Ispir, et~al\mbox{.}}{Cheng et~al\mbox{.}}{2016}]%
        {cheng2016wide}
\bibfield{author}{\bibinfo{person}{Heng-Tze Cheng}, \bibinfo{person}{Levent
  Koc}, \bibinfo{person}{Jeremiah Harmsen}, \bibinfo{person}{Tal Shaked},
  \bibinfo{person}{Tushar Chandra}, \bibinfo{person}{Hrishi Aradhye},
  \bibinfo{person}{Glen Anderson}, \bibinfo{person}{Greg Corrado},
  \bibinfo{person}{Wei Chai}, \bibinfo{person}{Mustafa Ispir}, {et~al\mbox{.}}}
  \bibinfo{year}{2016}\natexlab{}.
\newblock \showarticletitle{Wide \& Deep Learning for Recommender Systems}. In
  \bibinfo{booktitle}{{\em RecSys}}.
\newblock


\bibitem[\protect\citeauthoryear{Ebesu and Fang}{Ebesu and Fang}{2017}]%
        {Ebesu2017}
\bibfield{author}{\bibinfo{person}{Travis Ebesu} {and} \bibinfo{person}{Yi
  Fang}.} \bibinfo{year}{2017}\natexlab{}.
\newblock \showarticletitle{Neural Semantic Personalized Ranking for item
  cold-start recommendation}.
\newblock \bibinfo{journal}{{\em Information Retrieval Journal\/}}
  \bibinfo{volume}{20}, \bibinfo{number}{2}, \bibinfo{pages}{109--131}.
\newblock


\bibitem[\protect\citeauthoryear{Geng, Zhang, Bian, and Chua}{Geng
  et~al\mbox{.}}{2015}]%
        {geng2015learning}
\bibfield{author}{\bibinfo{person}{Xue Geng}, \bibinfo{person}{Hanwang Zhang},
  \bibinfo{person}{Jingwen Bian}, {and} \bibinfo{person}{Tat-Seng Chua}.}
  \bibinfo{year}{2015}\natexlab{}.
\newblock \showarticletitle{Learning image and user features for recommendation
  in social networks}. In \bibinfo{booktitle}{{\em ICCV}}.
\newblock


\bibitem[\protect\citeauthoryear{Gong and Zhang}{Gong and Zhang}{2016}]%
        {gong2016hashtag}
\bibfield{author}{\bibinfo{person}{Yuyun Gong} {and} \bibinfo{person}{Qi
  Zhang}.} \bibinfo{year}{2016}\natexlab{}.
\newblock \showarticletitle{Hashtag Recommendation Using Attention-Based
  Convolutional Neural Network.}. In \bibinfo{booktitle}{{\em IJCAI}}.
\newblock


\bibitem[\protect\citeauthoryear{Goodfellow, Bengio, and Courville}{Goodfellow
  et~al\mbox{.}}{2016}]%
        {GoodfellowDLBook}
\bibfield{author}{\bibinfo{person}{Ian Goodfellow}, \bibinfo{person}{Yoshua
  Bengio}, {and} \bibinfo{person}{Aaron Courville}.}
  \bibinfo{year}{2016}\natexlab{}.
\newblock \bibinfo{booktitle}{{\em Deep Learning}}.
\newblock \bibinfo{publisher}{MIT Press}.
\newblock
\newblock
\shownote{\url{http://www.deeplearningbook.org}.}


\bibitem[\protect\citeauthoryear{Graves, Wayne, Reynolds, Harley, Danihelka,
  Grabska-Barwi{\'n}ska, Colmenarejo, Grefenstette, Ramalho, Agapiou,
  et~al\mbox{.}}{Graves et~al\mbox{.}}{2016}]%
        {graves2016DiffNeuralComp}
\bibfield{author}{\bibinfo{person}{Alex Graves}, \bibinfo{person}{Greg Wayne},
  \bibinfo{person}{Malcolm Reynolds}, \bibinfo{person}{Tim Harley},
  \bibinfo{person}{Ivo Danihelka}, \bibinfo{person}{Agnieszka
  Grabska-Barwi{\'n}ska}, \bibinfo{person}{Sergio~G{\'o}mez Colmenarejo},
  \bibinfo{person}{Edward Grefenstette}, \bibinfo{person}{Tiago Ramalho},
  \bibinfo{person}{John Agapiou}, {et~al\mbox{.}}}
  \bibinfo{year}{2016}\natexlab{}.
\newblock \showarticletitle{Hybrid computing using a neural network with
  dynamic external memory}.
\newblock \bibinfo{journal}{{\em Nature\/}} \bibinfo{volume}{538},
  \bibinfo{number}{7626} (\bibinfo{year}{2016}), \bibinfo{pages}{471--476}.
\newblock


\bibitem[\protect\citeauthoryear{He, Zhang, Ren, and Sun}{He
  et~al\mbox{.}}{2015}]%
        {he2015delving}
\bibfield{author}{\bibinfo{person}{Kaiming He}, \bibinfo{person}{Xiangyu
  Zhang}, \bibinfo{person}{Shaoqing Ren}, {and} \bibinfo{person}{Jian Sun}.}
  \bibinfo{year}{2015}\natexlab{}.
\newblock \showarticletitle{Delving deep into rectifiers: Surpassing
  human-level performance on imagenet classification}. In
  \bibinfo{booktitle}{{\em CVPR}}.
\newblock


\bibitem[\protect\citeauthoryear{He, Zhang, Ren, and Sun}{He
  et~al\mbox{.}}{2016}]%
        {he2016resnet}
\bibfield{author}{\bibinfo{person}{Kaiming He}, \bibinfo{person}{Xiangyu
  Zhang}, \bibinfo{person}{Shaoqing Ren}, {and} \bibinfo{person}{Jian Sun}.}
  \bibinfo{year}{2016}\natexlab{}.
\newblock \showarticletitle{Deep residual learning for image recognition}. In
  \bibinfo{booktitle}{{\em CVPR}}.
\newblock


\bibitem[\protect\citeauthoryear{He, Liao, Zhang, Nie, Hu, and Chua}{He
  et~al\mbox{.}}{2017}]%
        {NeuralCF-2017}
\bibfield{author}{\bibinfo{person}{Xiangnan He}, \bibinfo{person}{Lizi Liao},
  \bibinfo{person}{Hanwang Zhang}, \bibinfo{person}{Liqiang Nie},
  \bibinfo{person}{Xia Hu}, {and} \bibinfo{person}{Tat-Seng Chua}.}
  \bibinfo{year}{2017}\natexlab{}.
\newblock \showarticletitle{Neural Collaborative Filtering}. In
  \bibinfo{booktitle}{{\em WWW}}.
\newblock


\bibitem[\protect\citeauthoryear{Hidasi, Karatzoglou, Baltrunas, and
  Tikk}{Hidasi et~al\mbox{.}}{2016}]%
        {hidasi2015session}
\bibfield{author}{\bibinfo{person}{Bal{\'a}zs Hidasi},
  \bibinfo{person}{Alexandros Karatzoglou}, \bibinfo{person}{Linas Baltrunas},
  {and} \bibinfo{person}{Domonkos Tikk}.} \bibinfo{year}{2016}\natexlab{}.
\newblock \showarticletitle{Session-based Recommendations with Recurrent Neural
  Networks}. In \bibinfo{booktitle}{{\em ICLR}}.
\newblock


\bibitem[\protect\citeauthoryear{Huang, Zhang, Gong, and Huang}{Huang
  et~al\mbox{.}}{2016}]%
        {huang2016hashtagMemoryNet}
\bibfield{author}{\bibinfo{person}{Haoran Huang}, \bibinfo{person}{Qi Zhang},
  \bibinfo{person}{Yeyun Gong}, {and} \bibinfo{person}{Xuanjing Huang}.}
  \bibinfo{year}{2016}\natexlab{}.
\newblock \showarticletitle{Hashtag Recommendation Using End-To-End Memory
  Networks with Hierarchical Attention.}. In \bibinfo{booktitle}{{\em COLING}}.
\newblock


\bibitem[\protect\citeauthoryear{Jannach and Ludewig}{Jannach and
  Ludewig}{2017}]%
        {Jannach2017SessionPlusNeighborhood}
\bibfield{author}{\bibinfo{person}{Dietmar Jannach} {and}
  \bibinfo{person}{Malte Ludewig}.} \bibinfo{year}{2017}\natexlab{}.
\newblock \showarticletitle{When Recurrent Neural Networks Meet the
  Neighborhood for Session-Based Recommendation}. In \bibinfo{booktitle}{{\em
  RecSys}}.
\newblock


\bibitem[\protect\citeauthoryear{Kabbur, Ning, and Karypis}{Kabbur
  et~al\mbox{.}}{2013}]%
        {kabbur2013fism}
\bibfield{author}{\bibinfo{person}{Santosh Kabbur}, \bibinfo{person}{Xia Ning},
  {and} \bibinfo{person}{George Karypis}.} \bibinfo{year}{2013}\natexlab{}.
\newblock \showarticletitle{Fism: factored item similarity models for top-n
  recommender systems}. In \bibinfo{booktitle}{{\em SIGKDD}}.
\newblock


\bibitem[\protect\citeauthoryear{Kim, Park, Oh, Lee, and Yu}{Kim
  et~al\mbox{.}}{2016}]%
        {Kim2016ConvMF}
\bibfield{author}{\bibinfo{person}{Donghyun Kim}, \bibinfo{person}{Chanyoung
  Park}, \bibinfo{person}{Jinoh Oh}, \bibinfo{person}{Sungyoung Lee}, {and}
  \bibinfo{person}{Hwanjo Yu}.} \bibinfo{year}{2016}\natexlab{}.
\newblock \showarticletitle{Convolutional Matrix Factorization for Document
  Context-Aware Recommendation}. In \bibinfo{booktitle}{{\em RecSys}}.
\newblock


\bibitem[\protect\citeauthoryear{Koren}{Koren}{2008}]%
        {koren2008factorization}
\bibfield{author}{\bibinfo{person}{Yehuda Koren}.}
  \bibinfo{year}{2008}\natexlab{}.
\newblock \showarticletitle{Factorization meets the neighborhood: a
  multifaceted collaborative filtering model}. In \bibinfo{booktitle}{{\em
  SIGKDD}}.
\newblock


\bibitem[\protect\citeauthoryear{Kumar, Irsoy, Ondruska, Iyyer, Bradbury,
  Gulrajani, Zhong, Paulus, and Socher}{Kumar et~al\mbox{.}}{2016}]%
        {Kumar2016AskMA}
\bibfield{author}{\bibinfo{person}{Ankit Kumar}, \bibinfo{person}{Ozan Irsoy},
  \bibinfo{person}{Peter Ondruska}, \bibinfo{person}{Mohit Iyyer},
  \bibinfo{person}{James Bradbury}, \bibinfo{person}{Ishaan Gulrajani},
  \bibinfo{person}{Victor Zhong}, \bibinfo{person}{Romain Paulus}, {and}
  \bibinfo{person}{Richard Socher}.} \bibinfo{year}{2016}\natexlab{}.
\newblock \showarticletitle{Ask Me Anything: Dynamic Memory Networks for
  Natural Language Processing}. In \bibinfo{booktitle}{{\em ICML}}.
\newblock


\bibitem[\protect\citeauthoryear{Li, Guo, and Mei}{Li et~al\mbox{.}}{2017}]%
        {Li2017MemNetForAttitudeId}
\bibfield{author}{\bibinfo{person}{Cheng Li}, \bibinfo{person}{Xiaoxiao Guo},
  {and} \bibinfo{person}{Qiaozhu Mei}.} \bibinfo{year}{2017}\natexlab{}.
\newblock \showarticletitle{Deep Memory Networks for Attitude Identification}.
  In \bibinfo{booktitle}{{\em WSDM}}.
\newblock


\bibitem[\protect\citeauthoryear{Li, Kawale, and Fu}{Li et~al\mbox{.}}{2015}]%
        {Li2015DCF}
\bibfield{author}{\bibinfo{person}{Sheng Li}, \bibinfo{person}{Jaya Kawale},
  {and} \bibinfo{person}{Yun Fu}.} \bibinfo{year}{2015}\natexlab{}.
\newblock \showarticletitle{Deep Collaborative Filtering via Marginalized
  Denoising Auto-encoder}. In \bibinfo{booktitle}{{\em CIKM}}.
\newblock


\bibitem[\protect\citeauthoryear{Linden, Smith, and York}{Linden
  et~al\mbox{.}}{2003}]%
        {Linden2003AmazonCF}
\bibfield{author}{\bibinfo{person}{Greg Linden}, \bibinfo{person}{Brent Smith},
  {and} \bibinfo{person}{Jeremy York}.} \bibinfo{year}{2003}\natexlab{}.
\newblock \showarticletitle{Amazon.Com Recommendations: Item-to-Item
  Collaborative Filtering}.
\newblock \bibinfo{journal}{{\em IEEE Internet Computing\/}}
  \bibinfo{volume}{7}, \bibinfo{number}{1}.
\newblock


\bibitem[\protect\citeauthoryear{Massa and Avesani}{Massa and Avesani}{2007}]%
        {MassaEpinionsDataset}
\bibfield{author}{\bibinfo{person}{Paolo Massa} {and} \bibinfo{person}{Paolo
  Avesani}.} \bibinfo{year}{2007}\natexlab{}.
\newblock \showarticletitle{Trust-aware Recommender Systems}. In
  \bibinfo{booktitle}{{\em RecSys}}.
\newblock


\bibitem[\protect\citeauthoryear{Ning and Karypis}{Ning and Karypis}{2011}]%
        {ning2011slim}
\bibfield{author}{\bibinfo{person}{Xia Ning} {and} \bibinfo{person}{George
  Karypis}.} \bibinfo{year}{2011}\natexlab{}.
\newblock \showarticletitle{Slim: Sparse linear methods for top-n recommender
  systems}. In \bibinfo{booktitle}{{\em ICDM}}.
\newblock


\bibitem[\protect\citeauthoryear{Rendle}{Rendle}{2010}]%
        {rendle2010factorization}
\bibfield{author}{\bibinfo{person}{Steffen Rendle}.}
  \bibinfo{year}{2010}\natexlab{}.
\newblock \showarticletitle{Factorization machines}. In
  \bibinfo{booktitle}{{\em ICDM}}.
\newblock


\bibitem[\protect\citeauthoryear{Rendle, Freudenthaler, Gantner, and
  Schmidt-thieme}{Rendle et~al\mbox{.}}{2009}]%
        {Rendle2009}
\bibfield{author}{\bibinfo{person}{Steffen Rendle}, \bibinfo{person}{Christoph
  Freudenthaler}, \bibinfo{person}{Zeno Gantner}, {and} \bibinfo{person}{Lars
  Schmidt-thieme}.} \bibinfo{year}{2009}\natexlab{}.
\newblock \showarticletitle{{BPR : Bayesian Personalized Ranking from Implicit
  Feedback}}.
\newblock \bibinfo{journal}{{\em UAI\/}} (\bibinfo{year}{2009}).
\newblock


\bibitem[\protect\citeauthoryear{Ricci, Rokach, and Shapira}{Ricci
  et~al\mbox{.}}{2011}]%
        {Ricci2011IntroRecSysHandbook}
\bibfield{author}{\bibinfo{person}{Francesco Ricci}, \bibinfo{person}{Lior
  Rokach}, {and} \bibinfo{person}{Bracha Shapira}.}
  \bibinfo{year}{2011}\natexlab{}.
\newblock \bibinfo{booktitle}{{\em Introduction to recommender systems
  handbook}}.
\newblock \bibinfo{publisher}{Springer}.
\newblock


\bibitem[\protect\citeauthoryear{Salakhutdinov, Mnih, and Hinton}{Salakhutdinov
  et~al\mbox{.}}{2007}]%
        {salakhutdinov2007restricted}
\bibfield{author}{\bibinfo{person}{Ruslan Salakhutdinov},
  \bibinfo{person}{Andriy Mnih}, {and} \bibinfo{person}{Geoffrey Hinton}.}
  \bibinfo{year}{2007}\natexlab{}.
\newblock \showarticletitle{Restricted Boltzmann machines for collaborative
  filtering}. In \bibinfo{booktitle}{{\em ICML}}.
\newblock


\bibitem[\protect\citeauthoryear{Sedhain, Menon, Sanner, and Xie}{Sedhain
  et~al\mbox{.}}{2015}]%
        {Sedhain2015}
\bibfield{author}{\bibinfo{person}{Suvash Sedhain},
  \bibinfo{person}{Aditya~Krishna Menon}, \bibinfo{person}{Scott Sanner}, {and}
  \bibinfo{person}{Lexing Xie}.} \bibinfo{year}{2015}\natexlab{}.
\newblock \showarticletitle{{AutoRec : Autoencoders Meet Collaborative
  Filtering}}.
\newblock \bibinfo{journal}{{\em WWW\/}}.
\newblock


\bibitem[\protect\citeauthoryear{Seo, Huang, Yang, and Liu}{Seo
  et~al\mbox{.}}{2017}]%
        {Seo2017ICN}
\bibfield{author}{\bibinfo{person}{Sungyong Seo}, \bibinfo{person}{Jing Huang},
  \bibinfo{person}{Hao Yang}, {and} \bibinfo{person}{Yan Liu}.}
  \bibinfo{year}{2017}\natexlab{}.
\newblock \showarticletitle{Interpretable Convolutional Neural Networks with
  Dual Local and Global Attention for Review Rating Prediction}. In
  \bibinfo{booktitle}{{\em RecSys}}.
\newblock


\bibitem[\protect\citeauthoryear{Sukhbaatar, Szlam, Weston, and
  Fergus}{Sukhbaatar et~al\mbox{.}}{2015}]%
        {Sukhbaatar2015EndToEndMN}
\bibfield{author}{\bibinfo{person}{Sainbayar Sukhbaatar},
  \bibinfo{person}{Arthur Szlam}, \bibinfo{person}{Jason Weston}, {and}
  \bibinfo{person}{Rob Fergus}.} \bibinfo{year}{2015}\natexlab{}.
\newblock \showarticletitle{End-To-End Memory Networks}. In
  \bibinfo{booktitle}{{\em NIPS}}.
\newblock


\bibitem[\protect\citeauthoryear{van~den Oord, Dieleman, and Schrauwen}{van~den
  Oord et~al\mbox{.}}{2013}]%
        {Oord2013DeepCM}
\bibfield{author}{\bibinfo{person}{A{\"{a}}ron van~den Oord},
  \bibinfo{person}{Sander Dieleman}, {and} \bibinfo{person}{Benjamin
  Schrauwen}.} \bibinfo{year}{2013}\natexlab{}.
\newblock \showarticletitle{{Deep content-based music recommendation}}.
\newblock \bibinfo{journal}{{\em NIPS\/}}.
\newblock


\bibitem[\protect\citeauthoryear{Wang and Blei}{Wang and Blei}{2011}]%
        {wang2011CTR}
\bibfield{author}{\bibinfo{person}{Chong Wang} {and} \bibinfo{person}{David~M
  Blei}.} \bibinfo{year}{2011}\natexlab{}.
\newblock \showarticletitle{Collaborative topic modeling for recommending
  scientific articles}. In \bibinfo{booktitle}{{\em SIGKDD}}.
\newblock


\bibitem[\protect\citeauthoryear{Wang, Wang, and Yeung}{Wang
  et~al\mbox{.}}{2015}]%
        {wang2015collaborative}
\bibfield{author}{\bibinfo{person}{Hao Wang}, \bibinfo{person}{Naiyan Wang},
  {and} \bibinfo{person}{Dit-Yan Yeung}.} \bibinfo{year}{2015}\natexlab{}.
\newblock \showarticletitle{Collaborative deep learning for recommender
  systems}. In \bibinfo{booktitle}{{\em SIGKDD}}.
\newblock


\bibitem[\protect\citeauthoryear{Wang, Yu, Zhang, Gong, Xu, Wang, Zhang, and
  Zhang}{Wang et~al\mbox{.}}{2017}]%
        {wang2017irgan}
\bibfield{author}{\bibinfo{person}{Jun Wang}, \bibinfo{person}{Lantao Yu},
  \bibinfo{person}{Weinan Zhang}, \bibinfo{person}{Yu Gong},
  \bibinfo{person}{Yinghui Xu}, \bibinfo{person}{Benyou Wang},
  \bibinfo{person}{Peng Zhang}, {and} \bibinfo{person}{Dell Zhang}.}
  \bibinfo{year}{2017}\natexlab{}.
\newblock \showarticletitle{IRGAN: A Minimax Game for Unifying Generative and
  Discriminative Information Retrieval Models}. In \bibinfo{booktitle}{{\em
  SIGIR}}.
\newblock


\bibitem[\protect\citeauthoryear{Weston, Chopra, and Bordes}{Weston
  et~al\mbox{.}}{2015}]%
        {Weston2014MemoryN}
\bibfield{author}{\bibinfo{person}{Jason Weston}, \bibinfo{person}{Sumit
  Chopra}, {and} \bibinfo{person}{Antoine Bordes}.}
  \bibinfo{year}{2015}\natexlab{}.
\newblock \showarticletitle{Memory Networks}. In \bibinfo{booktitle}{{\em
  ICLR}}.
\newblock


\bibitem[\protect\citeauthoryear{Wu, Ahmed, Beutel, Smola, and Jing}{Wu
  et~al\mbox{.}}{2017}]%
        {Wu2017RecurrentRecNet}
\bibfield{author}{\bibinfo{person}{Chao-Yuan Wu}, \bibinfo{person}{Amr Ahmed},
  \bibinfo{person}{Alex Beutel}, \bibinfo{person}{Alexander~J. Smola}, {and}
  \bibinfo{person}{How Jing}.} \bibinfo{year}{2017}\natexlab{}.
\newblock \showarticletitle{Recurrent Recommender Networks}. In
  \bibinfo{booktitle}{{\em WSDM}}.
\newblock


\bibitem[\protect\citeauthoryear{Wu, DuBois, Zheng, and Ester}{Wu
  et~al\mbox{.}}{2016}]%
        {Wu2016CDAE}
\bibfield{author}{\bibinfo{person}{Yao Wu}, \bibinfo{person}{Christopher
  DuBois}, \bibinfo{person}{Alice~X. Zheng}, {and} \bibinfo{person}{Martin
  Ester}.} \bibinfo{year}{2016}\natexlab{}.
\newblock \showarticletitle{Collaborative Denoising Auto-Encoders for Top-N
  Recommender Systems}. In \bibinfo{booktitle}{{\em WSDM}}.
\newblock


\bibitem[\protect\citeauthoryear{Xiao, Ye, He, Zhang, Wu, and Chua}{Xiao
  et~al\mbox{.}}{2017}]%
        {xiao2017attentional}
\bibfield{author}{\bibinfo{person}{Jun Xiao}, \bibinfo{person}{Hao Ye},
  \bibinfo{person}{Xiangnan He}, \bibinfo{person}{Hanwang Zhang},
  \bibinfo{person}{Fei Wu}, {and} \bibinfo{person}{Tat-Seng Chua}.}
  \bibinfo{year}{2017}\natexlab{}.
\newblock \showarticletitle{Attentional factorization machines: Learning the
  weight of feature interactions via attention networks}.
\newblock \bibinfo{journal}{{\em IJCAI\/}}.
\newblock


\bibitem[\protect\citeauthoryear{Xiong, Merity, and Socher}{Xiong
  et~al\mbox{.}}{2016}]%
        {xiong2016dynamic}
\bibfield{author}{\bibinfo{person}{Caiming Xiong}, \bibinfo{person}{Stephen
  Merity}, {and} \bibinfo{person}{Richard Socher}.}
  \bibinfo{year}{2016}\natexlab{}.
\newblock \showarticletitle{Dynamic memory networks for visual and textual
  question answering}. In \bibinfo{booktitle}{{\em ICML}}.
\newblock


\bibitem[\protect\citeauthoryear{Zhang, Yao, and Sun}{Zhang
  et~al\mbox{.}}{2017a}]%
        {zhang2017deep}
\bibfield{author}{\bibinfo{person}{Shuai Zhang}, \bibinfo{person}{Lina Yao},
  {and} \bibinfo{person}{Aixin Sun}.} \bibinfo{year}{2017}\natexlab{a}.
\newblock \showarticletitle{Deep Learning based Recommender System: A Survey
  and New Perspectives}.
\newblock \bibinfo{journal}{{\em arXiv:1707.07435\/}} (\bibinfo{year}{2017}).
\newblock


\bibitem[\protect\citeauthoryear{Zhang, Yao, and Xu}{Zhang
  et~al\mbox{.}}{2017b}]%
        {Zhang2017AutoSvd}
\bibfield{author}{\bibinfo{person}{Shuai Zhang}, \bibinfo{person}{Lina Yao},
  {and} \bibinfo{person}{Xiwei Xu}.} \bibinfo{year}{2017}\natexlab{b}.
\newblock \showarticletitle{AutoSVD++: An Efficient Hybrid Collaborative
  Filtering Model via Contractive Auto-encoders}. In \bibinfo{booktitle}{{\em
  SIGIR}}.
\newblock


\end{thebibliography}
\end{document}